\documentclass[preprint2]{aastex61}

\usepackage{graphicx}
\usepackage{lineno}                   
\usepackage{amsmath}                  
\usepackage{dcolumn}                  
\usepackage{bm}                       
\usepackage{xspace}                   
\usepackage{hyperref}                 
\usepackage{gensymb}

\graphicspath{{./}{figures/}}

\hypersetup{
  colorlinks=true,
  linkcolor=blue,
  citecolor=blue,
  urlcolor=blue,
}

\shorttitle{TeV gamma rays from passive Giant Molecular Clouds}
\shortauthors{HAWC Collaboration}

\begin{document}
\title{Probing the Sea of Cosmic Rays by Measuring Gamma-Ray Emission from Passive Giant Molecular Clouds with HAWC}

\author[0000-0003-0197-5646]{A.~Albert}
\affiliation{Physics Division, Los Alamos National Laboratory, Los Alamos, NM, USA }

\author[0000-0001-8749-1647]{R.~Alfaro}
\affiliation{Instituto de F\'{i}sica, Universidad Nacional Aut\'{o}noma de M\'{e}xico, Ciudad de M\'{e}xico, M\'{e}xico }

\author{C.~Alvarez}
\affiliation{Universidad Aut\'{o}noma de Chiapas, Tuxtla Guti\'{e}rrez, Chiapas, M\'{e}xico}

\author{J.R.~Angeles Camacho}
\affiliation{Instituto de F\'{i}sica, Universidad Nacional Aut\'{o}noma de M\'{e}xico, Ciudad de M\'{e}xico, M\'{e}xico }

\author{J.C.~Arteaga-Vel\'{a}zquez}
\affiliation{Universidad Michoacana de San Nicol\'{a}s de Hidalgo, Morelia, M\'{e}xico }

\author{K.P.~Arunbabu}
\affiliation{Instituto de Geof\'{i}sica, Universidad Nacional Aut\'{o}noma de M\'{e}xico, Ciudad de M\'{e}xico, M\'{e}xico }

\author{D.~Avila Rojas}
\affiliation{Instituto de F\'{i}sica, Universidad Nacional Aut\'{o}noma de M\'{e}xico, Ciudad de M\'{e}xico, M\'{e}xico }

\author[0000-0002-2084-5049]{H.A.~Ayala Solares}
\affiliation{Department of Physics, Pennsylvania State University, University Park, PA, USA }

\author{V.~Baghmanyan}
\affiliation{Institute of Nuclear Physics Polish Academy of Sciences, PL-31342 IFJ-PAN, Krakow, Poland}

\author[0000-0003-3207-105X]{E.~Belmont-Moreno}
\affiliation{Instituto de F\'{i}sica, Universidad Nacional Aut\'{o}noma de M\'{e}xico, Ciudad de M\'{e}xico, M\'{e}xico }

\author{S.Y.~BenZvi}
\affiliation{Department of Physics \& Astronomy, University of Rochester, Rochester, NY , USA}

\author[0000-0002-5493-6344]{C.~Brisbois}
\affiliation{Dept. of Physics, University of Maryland, College Park, MD 20742, USA}

\author[0000-0002-4042-3855]{K.S.~Caballero-Mora}
\affiliation{Universidad Aut\'{o}noma de Chiapas, Tuxtla Guti\'{e}rrez, Chiapas, M\'{e}xico}

\author[0000-0003-2158-2292]{T.~Capistr\'{a}n}
\affiliation{Instituto de Astronom\'{i}a, Universidad Nacional Aut\'{o}noma de M\'{e}xico, Ciudad de M\'{e}xico, M\'{e}xico }

\author[0000-0002-8553-3302]{A.~Carrami\~{n}ana}
\affiliation{Instituto Nacional de Astrof\'{i}sica, \'{O}ptica y Electr\'{o}nica, Puebla, M\'{e}xico }

\author[0000-0002-6144-9122]{S.~Casanova}
\affiliation{Institute of Nuclear Physics Polish Academy of Sciences, PL-31342 IFJ-PAN, Krakow, Poland }

\author[0000-0002-7607-9582]{U.~Cotti}
\affiliation{Universidad Michoacana de San Nicol\'{a}s de Hidalgo, Morelia, M\'{e}xico }

\author{J.~Cotzomi}
\affiliation{Facultad de Ciencias F\'{i}sico Matemáticas, Benem\'{e}rita Universidad Aut\'{o}noma de Puebla, Puebla, M\'{e}xico}

\author{S.~Couti\~{n}o de Le\'{o}n}
\affiliation{Instituto Nacional de Astrof\'{i}sica, \'{O}ptica y Electr\'{o}nica, Puebla, M\'{e}xico }

\author[0000-0001-9643-4134]{E.~De la Fuente}
\affiliation{Departamento de F\'{i}sica, Centro Universitario de Ciencias Exactas e Ingenierias, Universidad de Guadalajara, Guadalajara, M\'{e}xico }

\author{R.~Diaz Hernandez}
\affiliation{Instituto Nacional de Astrof\'{i}sica, \'{O}ptica y Electr\'{o}nica, Puebla, M\'{e}xico }

\author[0000-0001-8451-7450]{B.L.~Dingus}
\affiliation{Physics Division, Los Alamos National Laboratory, Los Alamos, NM, USA }

\author[0000-0002-2987-9691]{M.A.~DuVernois}
\affiliation{Department of Physics, University of Wisconsin-Madison, Madison, WI 53706, USA }

\author{M.~Durocher}
\affiliation{Physics Division, Los Alamos National Laboratory, Los Alamos, NM, USA }

\author[0000-0002-0087-0693]{J.C.~D\'{i}az-V\'{e}lez}
\affiliation{Departamento de F\'{i}sica, Centro Universitario de Ciencias Exactas e Ingenierias, Universidad de Guadalajara, Guadalajara, M\'{e}xico }

\author{R.W.~Ellsworth}
\affiliation{Dept. of Physics, University of Maryland, College Park, MD 20742, USA}

\author{K.~Engel}
\affiliation{Dept. of Physics, University of Maryland, College Park, MD 20742, USA}

\author[0000-0001-7074-1726]{C.~Espinoza}
\affiliation{Instituto de F\'{i}sica, Universidad Nacional Aut\'{o}noma de M\'{e}xico, Ciudad de M\'{e}xico, M\'{e}xico }

\author{K.L.~Fan}
\affiliation{Dept. of Physics, University of Maryland, College Park, MD 20742, USA}

\author{M.~Fern\'{a}ndez Alonso}
\affiliation{Department of Physics, Pennsylvania State University, University Park, PA, USA }


\author{N.~Fraija}
\affiliation{Instituto de Astronom\'{i}a, Universidad Nacional Aut\'{o}noma de M\'{e}xico, Ciudad de M\'{e}xico, M\'{e}xico }

\author{A.~Galv\'{a}n-G\'{a}mez}
\affiliation{Instituto de Astronom\'{i}a, Universidad Nacional Aut\'{o}noma de M\'{e}xico, Ciudad de M\'{e}xico, M\'{e}xico }

\author{D.~Garcia}
\affiliation{Instituto de F\'{i}sica, Universidad Nacional Aut\'{o}noma de M\'{e}xico, Ciudad de M\'{e}xico, M\'{e}xico }

\author[0000-0002-4188-5584]{J.A.~Garc\'{i}a-Gonz\'{a}lez}
\affiliation{Instituto de Astronom\'{i}a, Universidad Nacional Aut\'{o}noma de M\'{e}xico, Ciudad de M\'{e}xico, M\'{e}xico }

\author[0000-0003-1122-4168]{F.~Garfias}
\affiliation{Instituto de Astronom\'{i}a, Universidad Nacional Aut\'{o}noma de M\'{e}xico, Ciudad de M\'{e}xico, M\'{e}xico }

\author[0000-0002-5209-5641]{M.M.~Gonz\'{a}lez}
\affiliation{Instituto de Astronom\'{i}a, Universidad Nacional Aut\'{o}noma de M\'{e}xico, Ciudad de M\'{e}xico, M\'{e}xico }

\author[0000-0002-9790-1299]{J.A.~Goodman}
\affiliation{Dept. of Physics, University of Maryland, College Park, MD 20742, USA}

\author{J.P.~Harding}
\affiliation{Physics Division, Los Alamos National Laboratory, Los Alamos, NM, USA }

\author{S.~Hernandez}
\affiliation{Instituto de F\'{i}sica, Universidad Nacional Aut\'{o}noma de M\'{e}xico, Ciudad de M\'{e}xico, M\'{e}xico }

\author{B.~Hona}
\affiliation{Department of Physics and Astronomy, University of Utah, Salt Lake City, UT, USA }

\author[0000-0002-3808-4639]{D.~Huang}
\affiliation{Department of Physics, Michigan Technological University, Houghton, MI, USA }

\author[0000-0002-5527-7141]{F.~Hueyotl-Zahuantitla}
\affiliation{Universidad Aut\'{o}noma de Chiapas, Tuxtla Guti\'{e}rrez, Chiapas, M\'{e}xico}

\author{P.~H{\"u}ntemeyer}
\affiliation{Department of Physics, Michigan Technological University, Houghton, MI, USA }

\author[0000-0001-5811-5167]{A.~Iriarte}
\affiliation{Instituto de Astronom\'{i}a, Universidad Nacional Aut\'{o}noma de M\'{e}xico, Ciudad de M\'{e}xico, M\'{e}xico }


\author[0000-0003-4467-3621]{V.~Joshi}
\affiliation{Erlangen Centre for Astroparticle Physics, Friedrich-Alexander-Universit{\"a}t Erlangen-N{\"u}rnberg, D-91058 Erlangen, Germany}

\author{D.~Kieda}
\affiliation{Department of Physics and Astronomy, University of Utah, Salt Lake City, UT, USA }

\author{A.~Lara}
\affiliation{Instituto de Geof\'{i}sica, Universidad Nacional Aut\'{o}noma de M\'{e}xico, Ciudad de M\'{e}xico, M\'{e}xico }

\author{W.H.~Lee}
\affiliation{Instituto de Astronom\'{i}a, Universidad Nacional Aut\'{o}noma de M\'{e}xico, Ciudad de M\'{e}xico, M\'{e}xico }

\author{J.~Lee}
\affiliation{Natural Science Research Institute, University of Seoul, Seoul, Republic of Korea}

\author[0000-0001-5516-4975]{H.~Le\'{o}n Vargas}
\affiliation{Instituto de F\'{i}sica, Universidad Nacional Aut\'{o}noma de M\'{e}xico, Ciudad de M\'{e}xico, M\'{e}xico }

\author{J.T.~Linnemann}
\affiliation{Dept. of Physics and Astronomy, Michigan State University, East Lansing, MI 48824, USA}

\author[0000-0001-8825-3624]{A.L.~Longinotti}
\affiliation{Instituto Nacional de Astrof\'{i}sica, \'{O}ptica y Electr\'{o}nica, Puebla, M\'{e}xico }

\author[0000-0003-2810-4867]{G.~Luis-Raya}
\affiliation{Universidad Politecnica de Pachuca, Pachuca, Hgo, M\'{e}xico }

\author{J.~Lundeen}
\affiliation{Dept. of Physics and Astronomy, Michigan State University, East Lansing, MI 48824, USA}

\author[0000-0001-8088-400X]{K.~Malone}
\affiliation{Physics Division, Los Alamos National Laboratory, Los Alamos, NM, USA }

\author[0000-0001-9052-856X]{O.~Martinez}
\affiliation{Facultad de Ciencias F\'{i}sico Matemáticas, Benem\'{e}rita Universidad Aut\'{o}noma de Puebla, Puebla, M\'{e}xico }


\author{J.~Mart\'{i}nez-Castro}
\affiliation{Centro de Investigaci\'on en Computaci\'on, Instituto Polit\'ecnico Nacional, Mexico City, Mexico.}

\author[0000-0002-2610-863X]{J.A.~Matthews}
\affiliation{Dept of Physics and Astronomy, University of New Mexico, Albuquerque, NM, USA }

\author[0000-0002-8390-9011]{P.~Miranda-Romagnoli}
\affiliation{Universidad Aut\'{o}noma del Estado de Hidalgo, Pachuca, M\'{e}xico }

\author{J.A.~Morales-Soto}
\affiliation{Universidad Michoacana de San Nicol\'{a}s de Hidalgo, Morelia, M\'{e}xico }

\author[0000-0002-1114-2640]{E.~Moreno}
\affiliation{Facultad de Ciencias F\'{i}sico Matemáticas, Benem\'{e}rita Universidad Aut\'{o}noma de Puebla, Puebla, M\'{e}xico }

\author[0000-0002-7675-4656]{M.~Mostaf\'{a}}
\affiliation{Department of Physics, Pennsylvania State University, University Park, PA, USA }

\author{A.~Nayerhoda}
\affiliation{Institute of Nuclear Physics Polish Academy of Sciences, PL-31342 IFJ-PAN, Krakow, Poland }

\author[0000-0003-1059-8731]{L.~Nellen}
\affiliation{Instituto de Ciencias Nucleares, Universidad Nacional Aut\'{o}noma de M\'{e}xico, Ciudad de M\'{e}xico, M\'{e}xico }

\author[0000-0001-9428-7572]{M.~Newbold}
\affiliation{Department of Physics and Astronomy, University of Utah, Salt Lake City, UT, USA }

\author[0000-0002-6859-3944]{M.U.~Nisa}
\affiliation{Dept. of Physics and Astronomy, Michigan State University, East Lansing, MI 48824, USA}

\author[0000-0001-7099-108X]{R.~Noriega-Papaqui}
\affiliation{Universidad Aut\'{o}noma del Estado de Hidalgo, Pachuca, M\'{e}xico }

\author{N.~Omodei}
\affiliation{Department of Physics, Stanford University: Stanford, CA 94305–4060, USA}

\author{A.~Peisker}
\affiliation{Dept. of Physics and Astronomy, Michigan State University, East Lansing, MI 48824, USA}

\author{Y.~P\'{e}rez Araujo}
\affiliation{Instituto de Astronom\'{i}a, Universidad Nacional Aut\'{o}noma de M\'{e}xico, Ciudad de M\'{e}xico, M\'{e}xico }

\author[0000-0001-5998-4938]{E.G.~P\'{e}rez-P\'{e}rez}
\affiliation{Universidad Politecnica de Pachuca, Pachuca, Hgo, M\'{e}xico }

\author[0000-0002-6524-9769]{C.D.~Rho}
\affiliation{Natural Science Research Institute, University of Seoul, Seoul, Republic of Korea}

\author[0000-0003-1327-0838]{D.~Rosa-Gonz\'{a}lez}
\affiliation{Instituto Nacional de Astrof\'{i}sica, \'{O}ptica y Electr\'{o}nica, Puebla, M\'{e}xico }

\author{E.~Ruiz-Velasco}
\affiliation{Max-Planck Institute for Nuclear Physics, 69117 Heidelberg, Germany}


\author[0000-0002-8610-8703]{F.~Salesa Greus}
\affiliation{Institute of Nuclear Physics Polish Academy of Sciences, PL-31342 IFJ-PAN, Krakow, Poland }
\affiliation{Instituto de F\'{i}sica Corpuscular, CSIC, Universitat de Val\`{e}ncia, E-46980, Paterna, Valencia, Spain}

\author{A.~Sandoval}
\affiliation{Instituto de F\'{i}sica, Universidad Nacional Aut\'{o}noma de M\'{e}xico, Ciudad de M\'{e}xico, M\'{e}xico }

\author{M.~Schneider}
\affiliation{Dept. of Physics, University of Maryland, College Park, MD 20742, USA}

\author{J.~Serna-Franco}
\affiliation{Instituto de F\'{i}sica, Universidad Nacional Aut\'{o}noma de M\'{e}xico, Ciudad de M\'{e}xico, M\'{e}xico }

\author{A.J.~Smith}
\affiliation{Dept. of Physics, University of Maryland, College Park, MD 20742, USA}

\author[0000-0002-1492-0380]{R.W.~Springer}
\affiliation{Department of Physics and Astronomy, University of Utah, Salt Lake City, UT, USA }

\author{P.~Surajbali}
\affiliation{Max-Planck Institute for Nuclear Physics, 69117 Heidelberg, Germany}

\author{M.~Tanner}
\affiliation{Department of Physics, Pennsylvania State University, University Park, PA, USA }

\author[0000-0001-9725-1479]{K.~Tollefson}
\affiliation{Dept. of Physics and Astronomy, Michigan State University, East Lansing, MI 48824, USA}

\author[0000-0002-1689-3945]{I.~Torres}
\affiliation{Instituto Nacional de Astrof\'{i}sica, \'{O}ptica y Electr\'{o}nica, Puebla, M\'{e}xico }

\author{R.~Torres-Escobedo}
\affiliation{Departamento de F\'{i}sica, Centro Universitario de Ciencias Exactas e Ingenierias, Universidad de Guadalajara, Guadalajara, M\'{e}xico }

\author{R.~Turner}
\affiliation{Department of Physics, Michigan Technological University, Houghton, MI, USA }

\author{F.~Ure\~{n}a-Mena}
\affiliation{Instituto Nacional de Astrof\'{i}sica, \'{O}ptica y Electr\'{o}nica, Puebla, M\'{e}xico }

\author[0000-0001-6876-2800]{L.~Villase\~{n}or}
\affiliation{Facultad de Ciencias F\'{i}sico Matem\'{a}ticas, Benem\'{e}rita Universidad Aut\'{o}noma de Puebla, Puebla, M\'{e}xico }

\author{T.~Weisgarber}
\affiliation{Department of Chemistry and Physics, California University of Pennsylvania, California, Pennsylvania, USA}

\author{E.~Willox}
\affiliation{Dept. of Physics, University of Maryland, College Park, MD 20742, USA}


\author{H.~Zhou}
\affiliation{Tsung-Dao Lee Institute \& School of Physics and Astronomy, Shanghai Jiao Tong University, Shanghai, China}

\author{C.~de Le\'{o}n}
\affiliation{Universidad Michoacana de San Nicol\'{a}s de Hidalgo, Morelia, M\'{e}xico }

\collaboration{HAWC Collaboration}
\noaffiliation

\correspondingauthor{H.A.~Ayala Solares}
\email{hgayala@psu.edu}

\begin{abstract}
The study of high-energy gamma rays from passive Giant Molecular Clouds (GMCs) in our Galaxy is an indirect way to characterize and probe the paradigm of the ``sea'' of cosmic rays in distant parts of the Galaxy. By using data from the High Altitude Water Cherenkov (HAWC) observatory, we measure the gamma-ray flux above 1 TeV of a set of these clouds to test the paradigm.

We selected high-galactic latitude clouds that are in HAWC's field-of-view and which are within 1~kpc distance from the Sun. 
We find no significant excess emission in the cloud regions, nor when we perform a stacked log-likelihood analysis of GMCs. Using a Bayesian approach, we calculate 95\% credible intervals upper limits of the gamma-ray flux and estimate limits on the cosmic-ray energy density of these regions. These are the first limits to constrain gamma-ray emission in the  multi-TeV energy range ($>$1 TeV) using passive high-galactic latitude GMCs. Assuming that the main gamma-ray production mechanism is due to proton-proton interaction, the upper limits are consistent with a cosmic-ray flux and energy density similar to that measured at Earth.

\end{abstract}

\keywords{Astroparticle physics --- gamma rays: diffuse background --- cosmic rays --- Gamma-Ray Astronomy}
\section{Introduction}
\label{sec:Introduction}
Direct observations of cosmic rays measure the spectrum surrounding the vicinity of the solar system \citep{ams}. It is generally assumed that this spectrum is representative of the cosmic-ray flux across the Galaxy and it is usually referred as the ``sea" of cosmic rays. 
This assumption comes from the fact that cosmic rays, after leaving the acceleration regions, diffuse through the Galaxy due to the deflecting effect of the interstellar magnetic field to alter the path of charged particles. If this process occurs on a timescale of millions of years, their distribution becomes homogeneous and isotropic \citep{crProp}. 

A way to probe the paradigm of the ``sea" of cosmic rays is by using gamma-ray data. Since the beginning of high-energy gamma-ray astrophysics, it was recognized that the observation of gamma rays gives us the opportunity to measure the propagation and distribution of cosmic rays in distant parts of the Galaxy, since these can interact with the interstellar matter and radiation fields generating gamma-ray emission.
This has been suggested extensively in the literature by \cite{blackFazio, crtogammas, Aharonian2001, barometers}. The premise is to look for regions that act as targets for cosmic rays: clouds that are part of the interstellar medium. We focus on passive giant molecular clouds (GMCs), i.e. clouds with no cosmic-ray accelerators inside of them~\citep{crtogammas}. 
These clouds are large reservoirs of gas (mainly molecular hydrogen) and dust. Masses of these complexes are of the order of $\sim 10^5\, M_\odot$. Their average density is a hundred to a thousand times greater than the average density in the solar vicinity which is of the order of one to ten particles per cubic centimetre~\citep{ISM}. It is assumed that, due to due to the low probability of cosmic-ray sources inside the GMCs,  the main gamma-ray production mechanism in these sources is the decay of neutral pions, which are produced by the collision of high-energy cosmic rays with interstellar matter in the clouds. 

The flux of gamma rays is proportional to the flux of cosmic rays $\Phi_{CR}$, as well as the total mass $M$ of the GMC, and inversely proportional to the distance square $d^2$ of the GMC \citep[see for example][]{barometers}:

\begin{equation}\label{eq:modelGCR}
F_{\gamma} \propto\frac{M}{d^2}\Phi_{CR}.
\end{equation}

It is then important to know the masses and distances of the GMCs. 
Equation \ref{eq:modelGCR} has been used in other analyses using data from the \textit{Fermi}-LAT. 

The \textit{Fermi}-LAT collaboration presented their studies to understand the cosmic-ray propagation through gamma-ray data of the Orion molecular cloud~\citep{orion}, and of the molecular clouds Chamaleon, R Coronae Australis, Cepheus and Polaris~\citep{crISM}. In both publications the authors calculated gamma-ray emissivities and molecular mass conversion factors $X_{CO}$. 

Other studies using \textit{Fermi}-LAT data include the work by \cite{gcrI} and \cite{gcrII}, where they look at high-galactic latitude clouds that are part of the Gould Belt. They calculate the average cosmic-ray spectrum in the Galaxy.

Another example was presented in \cite{fermiCR}, where gamma-ray spectra above 300 MeV were used to extract the cosmic-ray spectra from eight massive clouds. 
They showed that the derived spectral indices and absolute fluxes of cosmic-ray protons in the energy interval 10~-~100 GeV agree with the direct measurements of local cosmic rays by the PAMELA experiment\citep{pamela}.

A similar study was published by \cite{seaCRFermi}. With their observations, they found that the flux of cosmic rays at distances from 0.6 kpc to 12.5 kpc also agrees with the locally measured cosmic-ray flux measured by the Alpha Magnetic Spectrometer (AMS) \citep{ams}. 

A recent publication by \cite{vardan}, also used \textit{Fermi}-LAT data to find discrepancies between the measured cosmic-ray flux at Earth and the cosmic-ray flux in distant regions of the Galaxy. They report that in three GMCs (Aquila Rift, Ophiuchi and Cepheus) they observe higher gamma-ray emission with respect to the expected emission from the ``sea of cosmic rays". Most of these publications have presented results of gamma rays up to 300 GeV.

While other TeV instruments have measured the cosmic-ray flux from GMCs from regions of the Galactic Plane \citep[e.g.][]{milagro,argoCyg,hessNat}, the present work is the first attempt to measure the ``sea of cosmic rays" using high-energy gamma rays above 1 TeV from high-latitude GMCs. In this paper we present measurements of gamma rays that have median energies of the order of $\sim$1-10~TeV using data from the High Altitude Water Cherenkov (HAWC) Observatory. 

The paper outline is as follows. We present the GMCs under study in Section \ref{sec:clouds}. The detector and the dataset are described in Section \ref{sec:Detector}. The analysis method is explained in Section \ref{sec:analysis}. The results and discussion are shown in Section \ref{sec:results}. Finally, we present our concluding remarks in Section \ref{sec:Conclusion}. 

\section{The Giant Molecular Clouds}
\label{sec:clouds}

Following  a similar procedure as the one presented in \cite{fermiCR}, we selected seven massive clouds identified by the CO Galactic survey of \cite{damesurvey} with the CfA 1.2 m millimetre-wave Telescope.
The selected clouds are Taurus, Orion, Perseus, Ophiuchi, Monoceros, Aquila Rift, and Hercules.  All of them are located less than 1\,kpc away from the solar system. These clouds have high galactic latitudes ($|b|>5^{\circ}$) and they are in the field of view of HAWC. Their masses and distances are listed in Table \ref{table:clouds}. The distances are obtained from \cite{damesurvey} and \cite{distances}. The masses are calculated using the Planck survey \citep{planck} (See appendix \ref{app:masses} for more details).

\begin{deluxetable*}{cccccccc}[!htb]
\tablewidth{6cm}
\tablecaption{Properties of the GMCs. Masses are calculated using the Planck survey \citep{planck} (See appendix \ref{app:masses} for mass calculations). Distances obtained from \cite{damesurvey} and \cite{distances} . The sky positions correspond to the center of the regions of interest that are analyzed. \label{table:clouds}}
\tablehead{
\colhead{ GMC } & \colhead{ Mass($\pm14\%$) [$10^5\, M_\odot$]} & \colhead{ Distance [pc]} & \colhead{A}&\colhead { l[$^{\circ}$]} & \colhead{b[$^{\circ}$]} & \colhead { Dec. [$^{\circ}$]} & \colhead{RA [$^{\circ}$]}}
\startdata
Taurus & 0.11 & 140$\pm$30 & 5.64 & 171.6 & -15.8  & 26.2 & 66.5 \\ 
Orion & 1.62  & 490$\pm$50 & 6.76 & -151.0 & -15.5  & -3.6 & 87.3 \\ 
Perseus & 0.15 & 315$\pm$32 & 1.55 & 159.0 & -20.5 & 30.9 & 52.9 \\ 
Ophiuchi & 0.06 & 125$\pm$18 & 3.94 & -5.7 & 16.8 & -23.5 & 247.5 \\ 
Monoceros & 0.6 & 830$\pm$83 & 0.87 & -145.8 & -12.4 & -6.8 & 92.3 \\ 
Aquila Rift & 0.77 & 225$\pm$55 & 15.27 & 22.5 & 11.8 & -3.72 & 267.6 \\ 
Hercules & 0.05 & 200$\pm$30 & 1.16 & 46.0 & 9.0 & 15.7 & 280.6 \\ 
\enddata
\end{deluxetable*}


The first four clouds that appear in Table \ref{table:clouds} (Taurus, Orion, Perseus and Ophiuchi) are considered as part of the Gould Belt according to \cite{gouldBelt}. The Gould Belt is a region in the Galaxy composed of gas and young stars. In the sky it appears as a ring with an inclination angle of $\sim20^{\circ}$ with respect to the galactic plane, although according to \cite{gouldBelt2}, it can be considered a disk-like structure. The clouds Aquila and Hercules do not line up with this structure and the Monoceros cloud is too far to be considered part of the belt.\footnote{A study of a Gould-belt-like structure in M83 makes comparisons including Monoceros as part of the Milky Way's Gould Belt~\citep{m83GB}. For the present paper, we will still consider Monoceros separately from the Gould Belt for the present paper.}  
However, a recent study by \cite{Radclwave}, has put into question the Gould Belt structure and substitutes it by the so called Radcliff Wave. In this case, the first three clouds in Table \ref{table:clouds} are part of the Radcliff Wave, while the fourth cloud is part of the Local Arm. 
Assuming that these clouds have similar properties, we combine the gamma-ray emission of the clouds that belong to the Gould Belt and the Radcliff Wave in a stacked analysis to increase the sensitivity of the probe\footnote{As it will be seen in $\S$ \ref{sec:results}, the results of the stacked clouds are similar due to the fact that Ophiuchi is at high zenith values for HAWC and hence HAWC has less sensitivity in this region.}. Due to the uncertainty of this classification, we also perform the stacked analysis for all the clouds. For completeness, we analyze each of them individually. 

\section{Observations and data analysis}
\subsection{The HAWC Data}
\label{sec:Detector}
The HAWC Observatory is a gamma-ray detector built in Sierra Negra in the Mexican state of Puebla at an altitude of 4100\,m above sea level. It is a wide field-of-view array of 300 water Cherenkov detectors (WCDs), with four photomultiplier tubes (PMTs) facing upwards in each WCD. The WCDs are cylindrical water tanks 4.5\,m high and 7.3\,m wide. The PMTs detect the Cherenkov light in the water from the passage of secondary particles, which are produced by gamma rays and cosmic rays interacting with the atmosphere. HAWC triggers with a rate of 25\,kHz and has a duty cycle of $>$95$\%$. More information on the HAWC observatory and the way air shower event data are reconstructed is presented in \cite{hawcCrab}.

For the analysis presented here, we use a dataset that started on 11/2014 and ended on 06/2019. The total integration time of the data is 1523 days.
As it was done in previous analyses, we apply gamma-hadron cuts to our dataset and divide the data in 9 bins. The data binning is done in the $f_{hit}$ parameter space, where $f_{hit}$ is defined by the ratio of the number of PMTs that were triggered by the air-shower event to the total number of active PMTs in the array. This is a proxy to an energy variable, where a lower $f_{hit}$ bin corresponds to a lower energy gamma-ray. 
The definition of the gamma-hadron cuts and the bounds of the $f_{hit}$ bins can be found in \cite{hawcCrab}.

\subsection{Analysis Method}
\label{sec:analysis}
The analysis is performed using the Multi-Mission Maximum Likelihood (3ML) \citep{threeml} framework together with the HAWC Accelerated Likelihood (HAL) plug-in\footnote{\url{https://github.com/threeML/hawc_hal}}. 
We perform both individual cloud and a stacked analyses, and assume the same spectral model for each cloud as a function of energy with the form 
\begin{equation}\label{eq:spectrum}
F_{\gamma}(E) = A\, C\, \left(\frac{E}{E_{0}}\right)^{-\alpha},
\end{equation}
where $\alpha$ is the spectral index, $E_0$ is the pivot energy, and $C$ is the normalization factor. The factor $A$, defined as $M_5/d^2_{\rm kpc}$ ($M_5=M/10^5 M_{\odot}$; $d_{\rm kpc} = d/1{\rm kpc}$) , expresses the gamma-ray visibility of a cloud of mass $M$, located at a distance $d$\citep{Aharonian2001, barometers}. This factor quantifies the weighted contribution of each cloud when we perform the stacking log-likelihood analysis. 

We fixed the spectral index to 2.7 assuming that gamma rays at these energies still follow the spectral shape of the ``sea'' of cosmic rays \citep{ams}. 
We also include morphology information from each cloud based on templates generated using the Planck survey (See Appendix \ref{app:templates} on how we built the templates). 
This information is combined, together with HAWC's detector response, to obtain an expected number of events in each $f_{\rm hit}$ bin used in the likelihood calculation.

For each cloud we create a log-likelihood profile as a function of $C$. For individual clouds, we use this profile to find the value $\hat{C}$ that optimizes the log-likelihood. For the stacking analysis, we add the log-likelihood profiles and then we find the value $\hat{C}$ that optimizes this profile. During the optimization we constrain the normalization value to be positive. 

A test statistic value $TS$ is calculated to check for a significant excess given the source model evaluated in the entire region of interest.
\begin{equation}\label{eq:ts}
TS = 2 \ln \left[\frac{\mathcal{L}(S(\hat{C})+B)}{\mathcal{L}(B)}\right],
\end{equation}
where S is the signal from the source model, while B is the background model, which corresponds to the background estimated from the data using direct integration~\citep[See ][]{hawcCat}. 

The best value $\hat{C}$ is then used as an input to a Markov-Chain Monte Carlo (MCMC) to estimate a probability distribution. Since we observe a non-significant detection (i.e. TS$<$25), we calculate the 95\% credible interval upper limits from the estimated distribution of the parameter $\hat{C}$. 

We compute a limit for the $E^{-2.7}$ spectrum model (refer afterwards as 2.7-model) and quasi-differential\footnote{The $f_{\rm hit}$ bins used in the analysis overlap when translated into energy space. Although this is taken into account in the analysis, the limits are therefore not strictly differential.} limits on the normalization in a similar way as previous HAWC analyses \cite{hawcFB, hawcDM}. The main difference between these two limits is that the quasi-differential limits are independent of any model assumption, while the limit for a broad energy range is model dependent. However, the quasi-differential limits are more conservative due to the lowering of the statistical power by dividing the data. 
For the 2.7-model limit, we use the energy band going above 1 TeV. For the quasi-differential limits, we define three energy half decade bands, between 1 TeV and 31.6 TeV and an overflow bin above 31.6 TeV. The range of the bands and the pivot energy, $E_0$, used in each range are shown in Table \ref{table:credint}. The pivot energy is set at 10 TeV for the 2.7-model limit.

\subsubsection{Significance Skymaps}
We produce significance skymaps of the GMC regions to complement the results. We use the same analysis method described in \cite{hawcCat2}. We calculate a likelihood ratio and then a test statistic value TS similar to Eq. \ref{eq:ts}. For the signal model, we assume a $E^{-2.7}$ spectrum and a fixed source morphology. The morphology used is a point source. 
If the null hypothesis is true, the significance distribution can be approximated as a gaussian distribution for the case of the point-source morphology. The maps and the distributions are shown in Fig \ref{fig:skymaps} in the Appendix \ref{app:results}.


\section{Results and Discussion}\label{sec:results}

We did not find significant emission from the studied regions (See tables in Appendix \ref{app:results}). Therefore, we proceed to calculate the 95\% credible intervals (C.I.) as explained in $\S$\ref{sec:analysis}. 
We also calculate the expected upper limits by performing the analysis with Poisson-fluctuated background-only maps. We obtain the 1- and 2- $\sigma$ confidence bands, as well as the median of the upper limits expected from a background-only scenario. 

\begin{figure*}[!htb]
    \figurenum{1}
    \centering
    
    \gridline{\fig{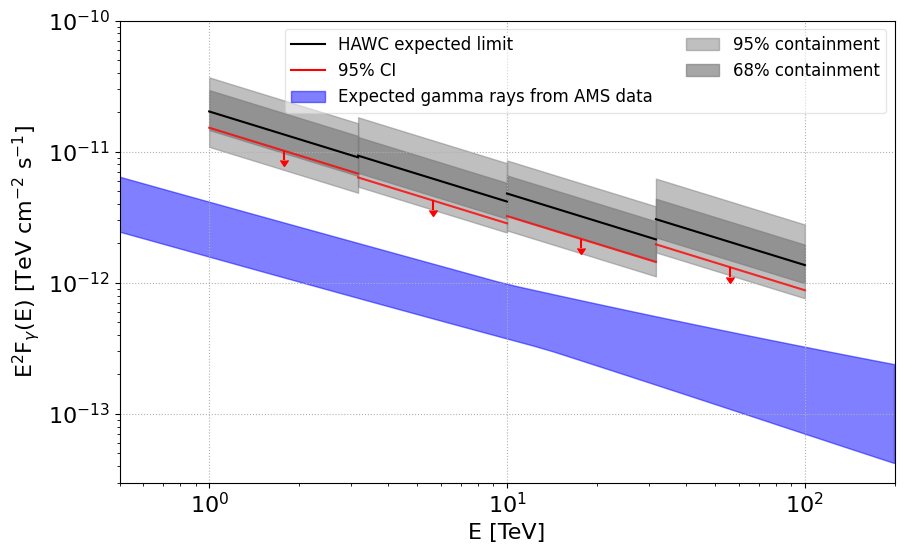}{0.35\textwidth}{(i) Aquila}        
              \fig{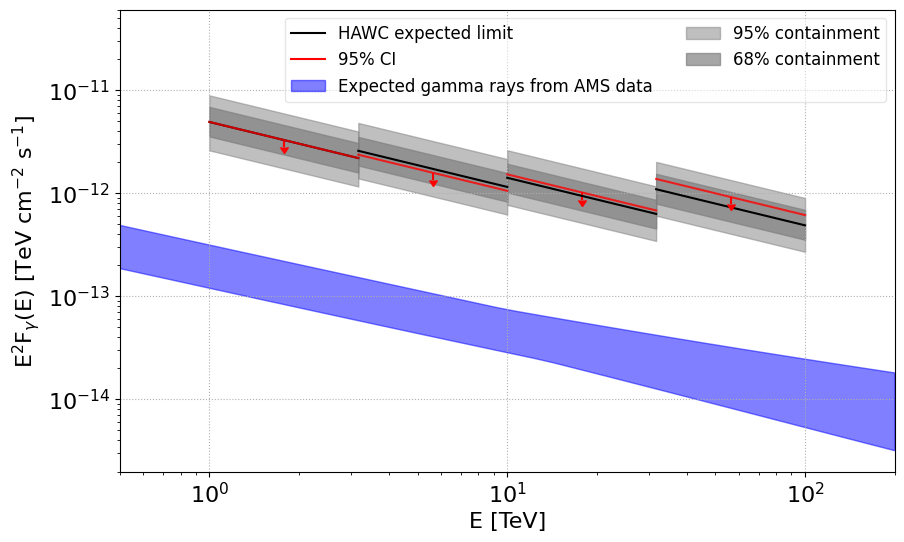}{0.35\textwidth}{(j) Hercules}}
    \caption{95\% C.I. upper limits on the gamma-ray flux of the giant molecular clouds studied. The gray band represents the statistical uncertainty in the U.L.(68\% and 90\% containment). The blue band is the expectation for the gamma-ray spectrum of the clouds based on local cosmic-ray measurements (Equation \ref{eq:gammaexp}). The width of the band corresponds to the use of two different software implementations to calculate the gamma flux~\citep{naima,crossSec}. (Continued)}
    \label{fig:uplims}
\end{figure*}

\begin{figure*}[!htb]
\centering
\fig{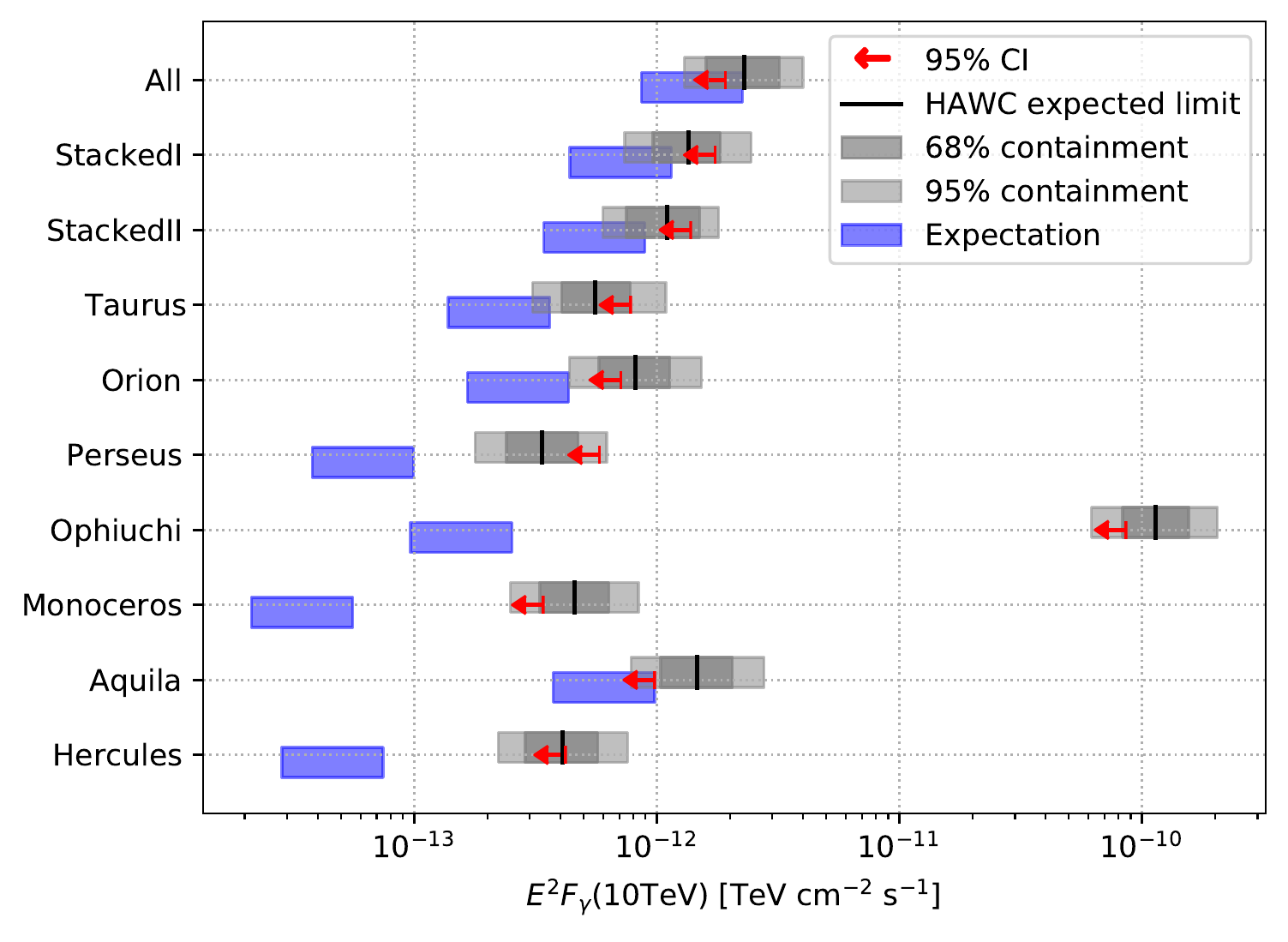}{0.6\textwidth}{}
\caption{95\% C.I. upper limits on the gamma-ray flux combining the energy bins from Figure \ref{fig:uplims}. The values plotted are at the pivot energy of 10 TeV and assuming an $E^{-2.7}$ spectrum. The gray and blue bands are calculated the same way as for Figure \ref{fig:uplims}. }
\label{fig:uplims2}
\end{figure*}

Figure \ref{fig:uplims} shows the quasi-differential upper limits. Figure \ref{fig:uplims2} shows the 2.7-model upper limits at 10 TeV.
These measurements can be compared with the gamma-ray flux expected from hadronic interactions producing gamma rays from neutral pion decay. The models were produced by using the parametrization of the $pp$ cross section ($\frac{d\sigma}{dE_{\gamma}}$) process from \cite{crossSec}. We use the cosmic-ray measurements from the AMS experiment \citep{ams} and extrapolate the AMS fit to higher cosmic-ray energies, with the assumption that it will maintain its spectral shape. 
We calculate the expected gamma-ray flux by convolving the cosmic-ray flux with the $pp$ cross section~\citep{seaCRFermi}\footnote{The software used to calculate the gamma-ray flux was Naima~\citep{naima} and libppgam~\citep{crossSec}. We show the results of both software packages as the blue band in Figures \ref{fig:uplims} and \ref{fig:uplims2} }:
\begin{equation}\label{eq:gammaexp}
    F_{\gamma}(E_{\gamma}) = \frac{\xi_N}{m_p} A \int dE_{p}\frac{d\sigma}{dE_{\gamma}} F_p(E_p),
\end{equation}
where $A$ is the same factor as equation \ref{eq:spectrum}, $\xi_N$ is the nuclear enhancement factor and assumed to be equal to 1.8 as  in \cite{seaCRFermi}, and $m_p$ is the proton mass.

As can be seen in Figure \ref{fig:uplims}, none of the limits are constraining the expected gamma-ray flux in the energy range accessible to HAWC. In Figure \ref{fig:uplims2} we see that if we stack all the clouds, the 2.7-model 95\% C.I. upper limit excludes part of the confidence interval of the gamma-ray emission expected from a cosmic-ray flux equal to that measured at Earth. However, there is not enough evidence to reject the paradigm of an unchanging sea of cosmic rays permeating our Galaxy. 
The weaker upper limit constraints measured in Ophiuchi are due to the fact that this region in the sky is at high zenith angles for HAWC, where the detector has lower sensitivity. None of the limits constrain the cosmic-ray flux to a lower value than the measured locally.  
Table \ref{table:credint} contains the observed and expected upper limits on the gamma-ray flux for each of the clouds and stacked analyses.
\begin{figure*}[htb!]
\centering
    \gridline{\fig{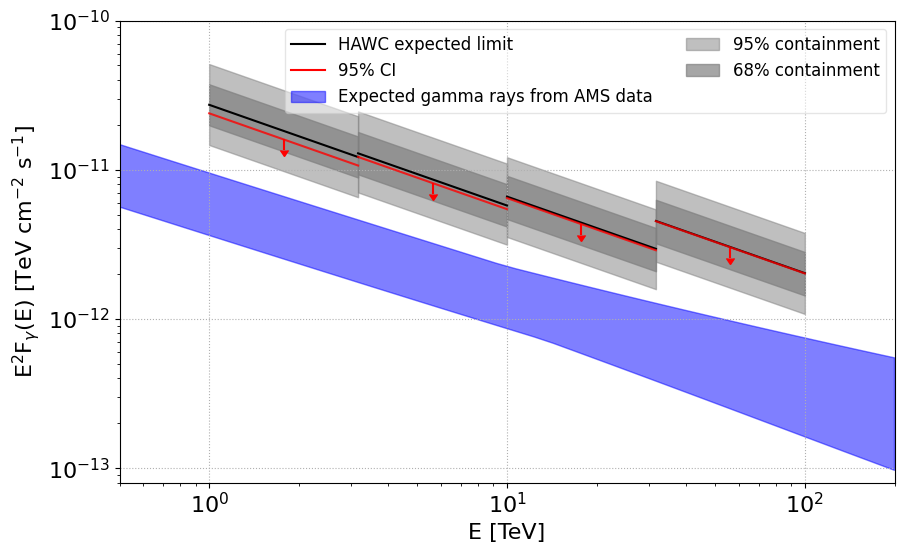}{0.35\textwidth}{(a) All}
        \fig{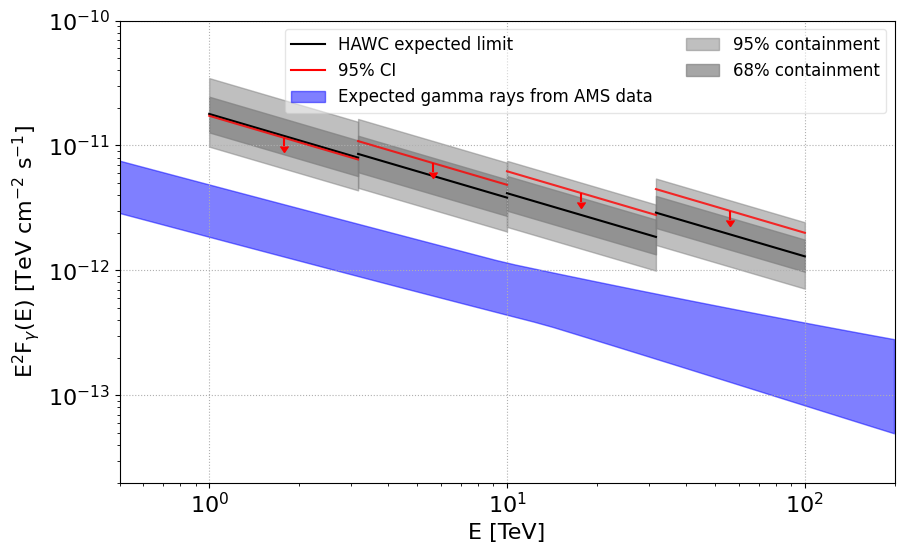}{0.35\textwidth}{(b) Gould Belt}}
    \gridline{\fig{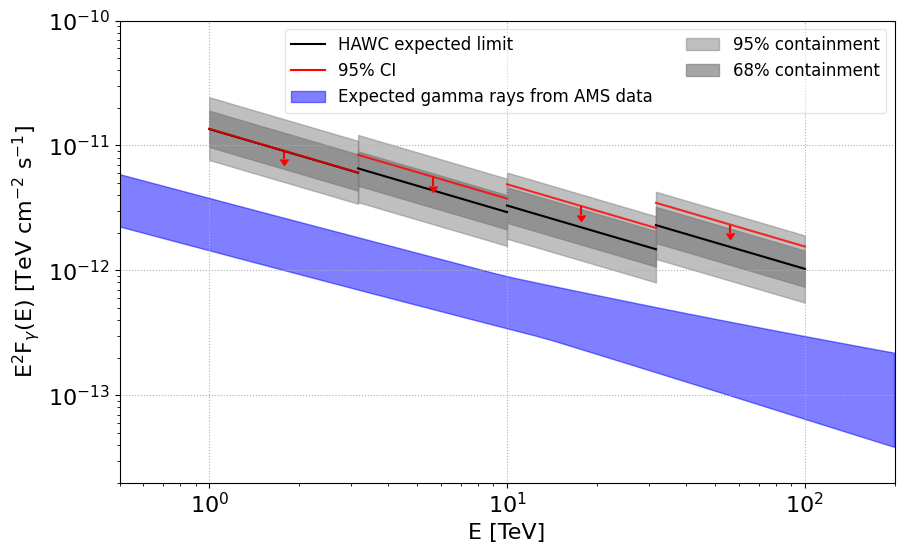}{0.35\textwidth}{(c) Radcliff}
        \fig{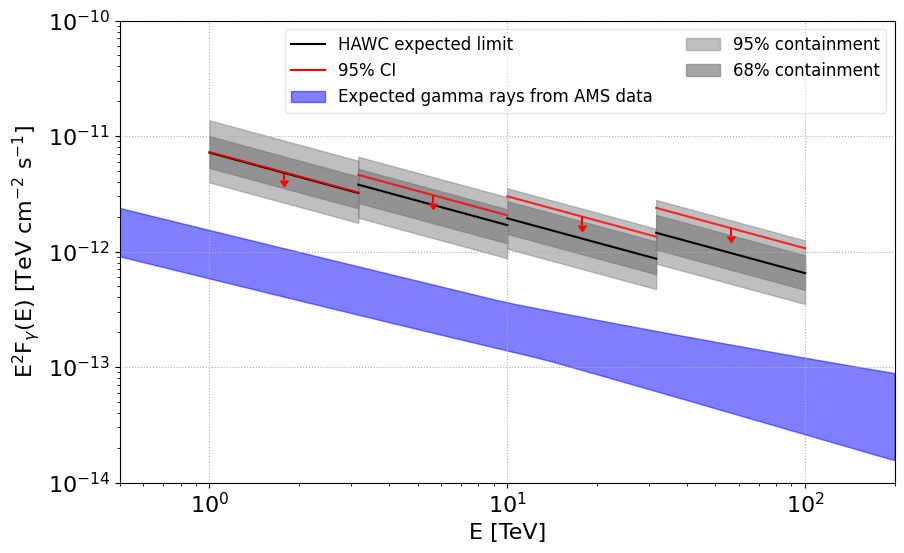}{0.35\textwidth}{(d) Taurus}}
    \gridline{\fig{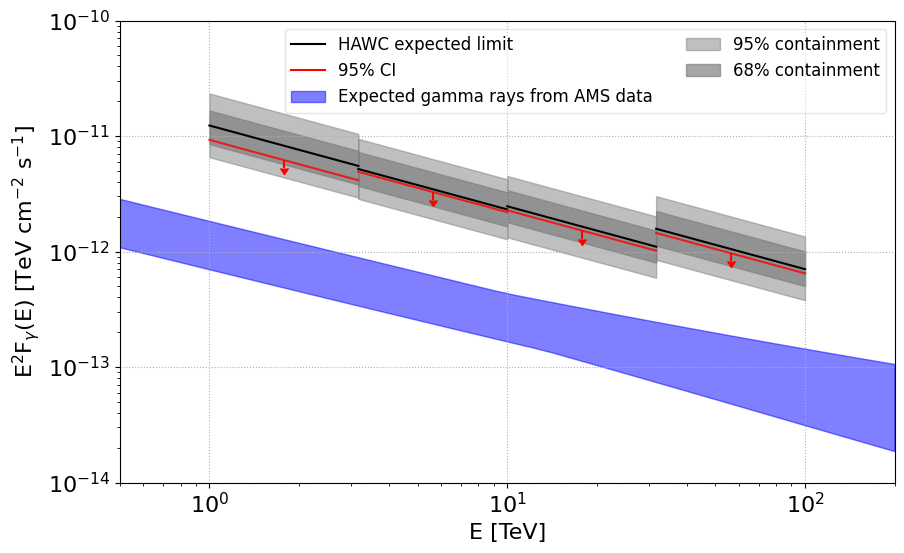}{0.35\textwidth}{(e) Orion}
              \fig{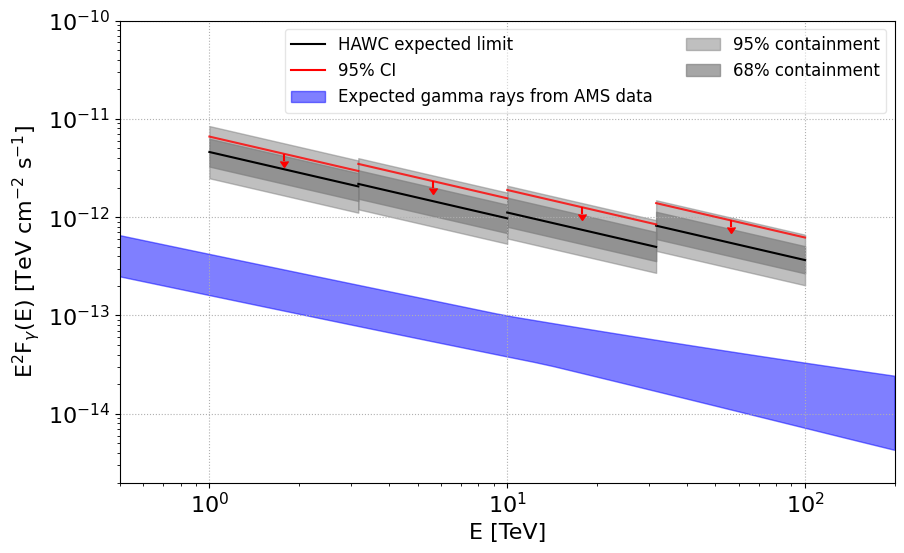}{0.35\textwidth}{(f) Perseus}}
    \gridline{\fig{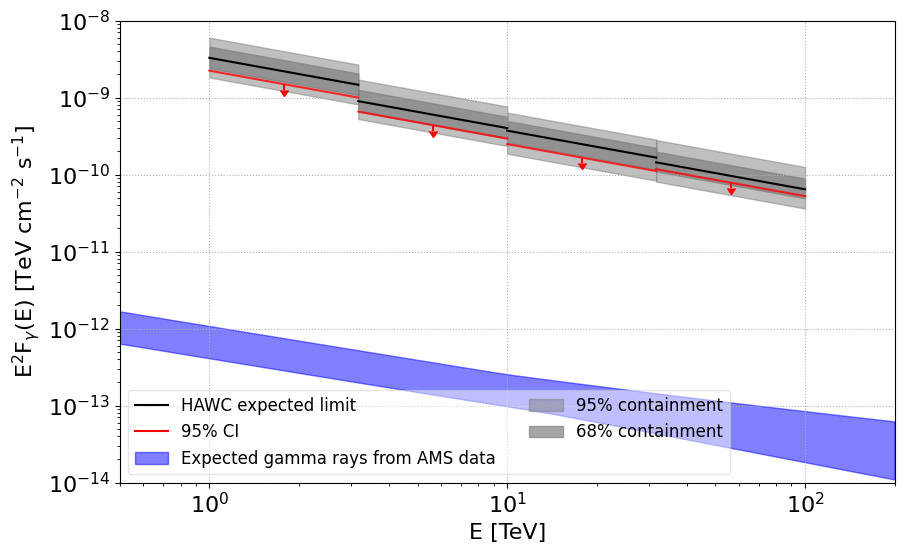}{0.35\textwidth}{(g) Ophiuchi}
              \fig{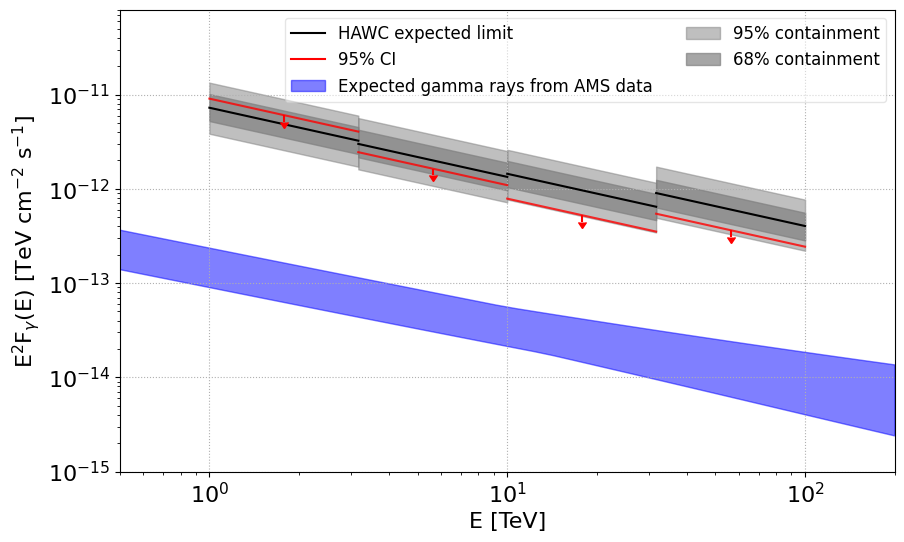}{0.35\textwidth}{(h) Monoceros}}
\setlength{\belowcaptionskip}{-20pt}
\caption{95\% C.I. upper limits on the gamma-ray flux of the giant molecular clouds studied. The gray band represents the statistical uncertainty in the U.L.(68$\%$ and 90$\%$ containment). The blue band is the expectation for the gamma-ray spectrum of the clouds based on local cosmic-ray measurements (Equation \ref{eq:gammaexp}). The width of the band corresponds to the use of two different software implementations to calculate the gamma flux~\citep{naima,crossSec}.}
\end{figure*}

\begin{longrotatetable}
\begin{deluxetable*}{ccccccccccc}
\tablecaption{Observed 95\% credible upper limits on the gamma-ray emission from the GMCs (and on $\hat{C}$), together with the expected median upper limits, as well as the 68\% and 95\% containment bands for the expected limits.\label{table:credint}}
\centering
\tablewidth{900pt}
\tabletypesize{\scriptsize}
\tablehead{
 \colhead{Energy Range} & \multicolumn{2}{c}{ 1 - 3.16 TeV } & \multicolumn{2}{c}{ 3.16 - 10.0 TeV } & \multicolumn{2}{c}{ 10.0 - 31.6 TeV } & \multicolumn{2}{c}{$>$31.6 TeV } & \multicolumn{2}{c}{ $>$1 TeV}\\
 \colhead{Pivot Energy} &\multicolumn{2}{c}{1.78 TeV}&\multicolumn{2}{c}{5.62 TeV} & \multicolumn{2}{c}{17.8 TeV} & \multicolumn{2}{c}{56.2 TeV} & \multicolumn{2}{c}{10 TeV}\\
  \colhead{} &\multicolumn{2}{c}{[$\times10^{-12}$]}&\multicolumn{2}{c}{[$\times10^{-13}$]} & \multicolumn{2}{c}{[$\times10^{-15}$]} & \multicolumn{2}{c}{[$\times10^{-16}$]} & \multicolumn{2}{c}{[$\times10^{-15}$]}\\
}
\startdata
  & U.L. & Expected Limit  &  U.L. & Expected Limit & U.L. & Expected Limit & U.L. & Expected Limit & U.L. & Expected Limit\\
  \textbf{All} & 5.0(0.1) & 5.8($_{-1.6}^{+2.1},\enspace_{-2.7}^{+5.0}$) & 2.6(0.1) & 2.7($_{-0.8}^{+1.1},\enspace_{-1.2}^{+2.5}$) & 13.6(0.4) & 14.0($_{-4.1}^{+5.3},\enspace_{-6.5}^{+12.0}$) & 9.5(0.3) & 9.6($_{-2.8}^{+3.7},\enspace_{-4.5}^{+8.2}$) & 19.0(0.5) & 23.0($_{-6.9}^{+8.8},\enspace_{-10.0}^{+16.0}$) \\
 \textbf{Gould Belt} & 3.6(0.2) & 3.7($_{-1.0}^{+1.5},\enspace_{-1.6}^{+3.6}$) & 2.3(0.1) & 1.8($_{-0.5}^{+0.7},\enspace_{-0.8}^{+1.1}$) & 13.5(0.8) & 8.7($_{-2.4}^{+3.3},\enspace_{-4.0}^{+7.3}$) & 9.4(0.7) & 6.1($_{-1.5}^{+2.2},\enspace_{-2.8}^{+5.4}$) & 18.0(1.0) & 13.5($_{-3.9}^{+4.8},\enspace_{-6.1}^{+11.0}$) \\
  \textbf{Radcliff} & 2.8(0.2) & 2.9($_{-0.8}^{+1.1},\enspace_{-1.3}^{+2.2}$) & 1.8(0.1) & 1.4($_{-0.4}^{+0.5},\enspace_{-0.7}^{+1.2}$)& 10.3(0.6) & 7.0($_{-1.9}^{+2.7},\enspace_{-3.2}^{+5.8}$) & 7.4(0.5) & 4.9($_{-1.4}^{+1.9},\enspace_{-2.3}^{+4.1}$) & 13.8(1.0) & 11.0($_{-3.5}^{+4.0},\enspace_{-5.0}^{+7.0}$) \\
 \textbf{Taurus} & 1.5(0.3) & 1.5($_{-0.4}^{+0.6},\enspace_{-0.7}^{+1.4}$) & 1.0(0.2) & 0.8($_{-0.2}^{+0.3},\enspace_{-0.4}^{+0.6}$) & 6.4(1.1) & 4.1($_{-1.1}^{+1.7},\enspace_{-1.9}^{+3.4}$) & 5.0(0.9) & 3.1($_{-0.9}^{+1.3},\enspace_{-1.4}^{+2.8}$) & 7.8(1.4) & 5.6($_{-1.6}^{+2.2},\enspace_{-2.6}^{+5.3}$) \\
 \textbf{Orion} & 2.0(0.3) & 2.6($_{-0.8}^{+0.9},\enspace_{-1.2}^{+2.3}$) & 1.0(0.1) & 1.0($_{-0.2}^{+0.5},\enspace_{-0.4}^{+1.0}$) & 4.8(0.7) & 5.2($_{-1.4}^{+2.0},\enspace_{-2.4}^{+4.3}$) & 3.0(0.4) & 3.3($_{-0.9}^{+1.4},\enspace_{-1.5}^{+3.1}$) & 7.1(1.0) & 8.2($_{-2.4}^{+3.1},\enspace_{-3.8}^{+7.1}$)\\
 \textbf{Perseus} & 1.4(0.9) & 1.0($_{-0.3}^{+0.3},\enspace_{-0.5}^{+0.8}$) & 0.74(0.5) & 0.5($_{-0.2}^{+0.1},\enspace_{-0.3}^{+0.3}$) & 4.0(2.6) & 2.4($_{-0.7}^{+1.1},\enspace_{-1.1}^{+2.0}$) & 2.9(1.9) & 1.7($_{-0.4}^{+0.7},\enspace_{-0.7}^{+1.4}$) & 5.8(3.7) & 3.4($_{-1.0}^{+1.3},\enspace_{-1.6}^{+2.8}$)\\
 \textbf{Ophiuchi [x100]} & 4.7(1.2) & 6.9($_{-1.9}^{+2.8},\enspace_{-3.1}^{+6.1}$) & 1.4(0.4) & 1.9($_{-0.5}^{+0.8},\enspace_{-0.8}^{+1.7}$) & 5.2(1.3) & 7.8($_{-2.3}^{+2.2},\enspace_{-3.9}^{+5.2}$) & 2.5(0.6) & 3.0($_{-0.7}^{+1.2},\enspace_{-1.3}^{+2.9}$ & 8.6(2.2) & 11.0($_{-2.7}^{+5.0},\enspace_{-4.8}^{+9.0}$)\\
 \textbf{Monoceros} & 1.9(2.2) & 1.5($_{-0.4}^{+0.6},\enspace_{-0.7}^{+1.3}$) & 0.5(0.6) & 0.6($_{-0.1}^{+0.3},\enspace_{-0.3}^{+0.6}$) & 1.7(2.0) & 3.1($_{-0.9}^{+1.1},\enspace_{-1.5}^{+2.4}$) & 1.2(1.4) & 1.9($_{-0.6}^{+0.7},\enspace_{-0.9}^{+1.7}$) & 3.4(3.9) & 4.6($_{-1.3}^{+1.7},\enspace_{-2.1}^{+3.8}$) \\
\textbf{Aquila} & 3.2(0.2) & 4.3($_{-1.2}^{+1.9},\enspace_{-2.0}^{+3.5}$) & 1.3(0.1) & 2.0($_{-0.5}^{+0.7},\enspace_{-0.9}^{+1.9}$) & 6.9(0.5) & 10.1($_{-3.2}^{+3.8},\enspace_{-4.8}^{+8.0}$) & 4.2(0.3) & 6.5($_{-1.8}^{+2.7},\enspace_{-2.9}^{+6.7}$) & 9.8(0.6) & 14.7($_{-4.3}^{+5.8},\enspace_{-6.9}^{+13.0}$)\\
 \textbf{Hercules} & 1.0(0.9) & 1.0($_{-0.3}^{+0.4},\enspace_{-0.5}^{+0.9}$) & 0.5(0.4) & 0.5($_{-0.1}^{+0.2},\enspace_{-0.2}^{+0.5}$) & 3.2(2.8) & 3.0($_{-0.9}^{+1.1},\enspace_{-1.4}^{+2.5}$) & 2.9(2.5) & 2.3($_{-0.6}^{+0.9},\enspace_{-1.0}^{+1.9}$) & 4.2(3.6) & 4.1($_{-1.2}^{+1.6},\enspace_{-1.9}^{+3.5}$) \\
\enddata
\tablenotetext{}{\textbf{Note. }The upper limits on the normalization $\hat{C}$ in parenthesis, are obtained by dividing the gamma-ray flux upper limits by the corresponding factor A. Flux units are in TeV$^{-1}$ cm$^{-2}$ s$^{-1}$.}
\tablenotetext{}{\textbf{Note. }Gould Belt includes Taurus, Orion, Perseus, and Ophiuchi; Radcliff includes Taurus, Orion, and Perseus. Notice that the flux values of Ophiuchi are a hundred times higher.}
\end{deluxetable*}
\end{longrotatetable}

\subsection{Systematic uncertainties}

\subsubsection{Detector systematic uncertainties}
We quantify systematic uncertainties by re-running the analysis with a range of different detector response files corresponding to our best understanding of PMT performance, detector calibration, and uncertainties in the point-spread function~\citep{hawcCrab}, resulting in a systematic relative uncertainty ranging from -15\% to 45\%. This means that the upper limits in Table \ref{table:credint} go up or down by these quantities. 

\subsubsection{Systematic uncertainty on the A factor}
The systematic uncertainty on the A factor depends only on the uncertainty on the mass of the cloud, since the distance cancels out when calculating A. The uncertainty on the mass is 14\%. The propagation of this uncertainty into the gamma-ray flux limits gives a systematic uncertainty between ranging from -8\% and 3\%.

\subsubsection{Systematic uncertainties due to spectral index assumptions}
Although our election for the power-law index is motivated from the cosmic-ray spectrum, we quantify how much our limits would change if we assume another value for the index. We use indexes of 2.5 and 2.9 ($\pm0.2$ from 2.7) for this. 
For the quasi-differential limits the systematic uncertainties due to spectral assumption range from -6\% to 6\%.
In the case of the model limit, we found that the value of the limit at the pivot energy ranges from -9\% to 14\%. 
Considering the uncertainties in the model limit, we calculate the upper limit on the integral flux above 1 TeV of the regions using the normalization value from the last column in Table \ref{table:credint}. The results of this systematic study are shown in Table \ref{table:intflux}. 

\begin{deluxetable}{cccc}
\tablecaption{Upper limits on the integral gamma-ray flux for 3 different spectral models. \label{table:intflux}}
\tablehead{
\colhead{} & {2.7-model} & {2.5-model} & {2.9-model} \\}
\startdata
\textbf{All} & 5.7 & 4.2 & 8.6 \\
\textbf{Gould Belt} & 5.3 & 4.0 & 6.3 \\
\textbf{Radcliff} & 4.1 & 3.2 & 5.1 \\
\textbf{Taurus} & 2.3 & 1.8 & 2.8 \\
\textbf{Orion} & 2.1 & 1.5 & 2.6 \\
\textbf{Perseus} & 1.7 & 1.3 & 2.1 \\
\textbf{Ophiuchi[x100]} & 2.5 & 1.6 & 3.5 \\
\textbf{Monoceros} & 1.1 & 0.7 & 1.5 \\
\textbf{Aquila} & 2.9 & 2.0 & 3.7 \\
\textbf{Hercules} & 1.2 & 0.9 & 1.6 \\
\enddata
\tablenotetext{}{\textbf{Note. } Flux units are in $\times 10^{-12}$ cm$^{-2}$ s$^{-1}$.}
\end{deluxetable}

\subsection{Constraints on the cosmic-ray energy density}

With the assumption that the gamma-ray emission is produced by pion decay, we estimate constraints in the cosmic-ray energy density from these distant regions and compare it to that measured in the local neighborhood. Using the same expression as in \cite{hessNat}, and using the same energy bins as in Table \ref{table:credint}, we can modify our spectral model, Equation \ref{eq:spectrum}, so that our free parameter is the cosmic-ray energy density (in $\text{eV}/\text{cm}^3$) 
\begin{eqnarray}\label{eq:crdensity}
    \rho_{CR} = 1.8 \times 10^{-2}\, &\left( \frac{\xi_{N}}{1.5}\right)^{-1} \left(\frac{L_{\gamma}}{10^{34}{\rm erg/s}}\right) \left( \frac{M}{10^6 M_{\odot}}\right)^{-1}, \nonumber \\
\end{eqnarray}
where $\xi_N$ accounts for nuclei heavier than hydrogen in both cosmic rays and interstellar matter, $L_{\gamma}$ is the luminosity of gamma rays, and $M$ is the mass of the region. 
The luminosity is obtained as
\begin{equation}\label{eq:luminosity}
L_{\gamma} (E_0 <E_{\gamma} <E_f) = 4 \pi d^2 \int^{E_f}_{E_0} E_{\gamma} F(E_{\gamma}) dE_{\gamma},
\end{equation}
where $d$ is the distance to the region.
We then insert Equation \ref{eq:luminosity} into Equation \ref{eq:crdensity}. After doing the algebra and taking care of units, we obtain
\begin{equation}\label{eq:crconst}
    \rho_{CR} = 4.930 \, C \, E_0 ^{\alpha} \, (E_i^{-\alpha + 2} - E_f^{-\alpha +2 }),
\end{equation}
where $\alpha$ is the index, $E_0$ is the mid-point of the energy bin and $E_i$, $E_f$ are the edges of the energy bins.
Equation~\ref{eq:spectrum} now can be rewritten so that the free parameter is $\rho_{CR}$. We apply the procedure in $\S$\ref{sec:analysis} to get an estimate of the average of the cosmic-ray energy density in the local Galaxy. 

\begin{deluxetable*}{ccccccccc}
\tablewidth{6cm}
\tablecaption{Observed 95\% credible upper limits on the cosmic-ray energy density from the GMCs, together with the expected median upper limits, as well as the 68\% and 95\% containment bands for the expected limits.\label{table:CRcredint}}
\tablehead{
 \colhead{Energy Range} & \multicolumn{2}{c}{ 1 - 3.16 TeV } & \multicolumn{2}{c}{ 3.16 - 10.0 TeV } & \multicolumn{2}{c}{ 10.0 - 31.6 TeV } & \multicolumn{2}{c}{$>$31.6 TeV } \\
 \colhead{Pivot Energy} &\multicolumn{2}{c}{1.78 TeV}&\multicolumn{2}{c}{5.62 TeV} & \multicolumn{2}{c}{17.8 TeV} & \multicolumn{2}{c}{56.2 TeV} \\
}
\startdata
  & U.L. & Expected Limit  &  U.L. & Expected Limit & U.L. & Expected Limit & U.L. & Expected Limit\\
\textbf{All} & 1.9 & 1.1($_{-0.3}^{+0.5},\enspace_{-0.6}^{+1.0}$) & 0.8 & 0.5($_{-0.1}^{+0.2},\enspace_{-0.2}^{+0.4}$) & 0.4 & 0.2($_{-0.07}^{+0.1},\enspace_{-0.1}^{+0.2}$) & 0.2 & 0.1($_{-0.03}^{+0.05},\enspace_{-0.07}^{+0.1}$) \\
\textbf{Gould Belt} & 2.7 & 2.7($_{-0.8}^{+1.0},\enspace_{-1.3}^{+2.0}$) & 1.8 & 1.4($_{-0.4}^{+0.6},\enspace_{-0.6}^{+1.2}$) & 1.0 & 0.8($_{-0.3}^{+0.3},\enspace_{-0.4}^{+1.0}$) & 0.7 & 0.5($_{-0.1}^{+0.2},\enspace_{-0.2}^{+0.6}$)  \\
\textbf{Radcliff} & 2.7 & 2.7($_{-0.7}^{+1.2},\enspace_{-1.2}^{+2.0}$) & 1.5 & 1.3($_{-0.4}^{+0.4},\enspace_{-0.6}^{+1.0}$) & 0.9 & 0.7($_{-0.2}^{+0.3},\enspace_{-0.3}^{+0.7}$) & 0.9 & 0.5($_{-0.2}^{+0.2},\enspace_{-0.3}^{+0.5}$) \\
\textbf{Taurus} & 3.6 & 3.5($_{-1.0}^{+1.3},\enspace_{-1.7}^{+3.2}$) & 2.2 & 1.8($_{-0.5}^{+0.6},\enspace_{-0.8}^{+1.6}$) & 1.4 & 1.0($_{-0.3}^{+0.4},\enspace_{-0.5}^{+0.7}$) & 1.2 & 0.7($_{-0.2}^{+0.3},\enspace_{-0.3}^{+0.6}$)  \\
\textbf{Orion} & 3.7 & 5.0($_{-1.5}^{+1.8},\enspace_{-2.3}^{+3.9}$) & 2.0 & 2.1($_{-0.6}^{+0.8},\enspace_{-0.9}^{+1.7}$) & 0.9 & 1.0($_{-0.3}^{+0.4},\enspace_{-0.5}^{+0.8}$) & 0.6 & 0.7($_{-0.2}^{+0.3},\enspace_{-0.3}^{+0.5}$)  \\
\textbf{Perseus} & 11.3 & 8.3($_{-2.4}^{+2.9},\enspace_{-3.9}^{+6.6}$) & 6.0 & 4.0($_{-1.0}^{+1.5},\enspace_{-1.8}^{+3.3}$) & 3.4 & 2.2($_{-0.6}^{+0.7},\enspace_{-1.0}^{+1.8}$) & 2.4 & 1.6($_{-0.4}^{+0.5},\enspace_{-0.7}^{+1.2}$)  \\
\textbf{Ophiuchi [x1000]} & 1.6  & 2.5($_{-0.8}^{+0.8},\enspace_{-1.3}^{+2.5}$) &  0.5 & 0.6($_{-0.2}^{+0.2},\enspace_{-0.3}^{+0.5}$) & 0.2 & 0.2($_{-0.06}^{+0.08},\enspace_{-0.1}^{+0.2}$) & 0.08 & 0.1($_{-0.02}^{+0.03},\enspace_{-0.03}^{+0.08}$) \\
\textbf{Monoceros} & 28.6 & 24.2($_{-6.5}^{+8.3},\enspace_{-11.6}^{+26.8}$) & 7.9 & 8.4($_{-2.2}^{+3.2},\enspace_{-3.6}^{+6.0}$) & 2.4 & 3.9($_{-1.1}^{+1.8},\enspace_{-1.7}^{+3.5}$) & 2.0 & 2.6($_{-0.7}^{+1.1},\enspace_{-1.2}^{+2.8}$)  \\
\textbf{Aquila} & 2.8 & 3.7($_{-1.1}^{+1.4},\enspace_{-1.7}^{+3.1}$) & 1.2 & 1.8($_{-0.5}^{+0.6},\enspace_{-0.8}^{+1.3}$) & 0.6 & 0.9($_{-0.2}^{+0.3},\enspace_{-0.4}^{+0.7}$) & 0.4 & 0.6($_{-0.2}^{+0.2},\enspace_{-0.3}^{+0.5}$)  \\
\textbf{Hercules} & 11.8 & 12.0($_{-3.2}^{+4.5},\enspace_{-5.3}^{+12.8}$) & 5.6 & 6.2($_{-1.8}^{+2.4},\enspace_{-2.9}^{+5.0}$) & 3.7 & 3.2($_{-0.9}^{+1.4},\enspace_{-1.4}^{+3.0}$) & 3.2 & 2.5($_{-0.7}^{+1.0},\enspace_{-1.1}^{+2.2}$)  \\
\enddata
\tablenotetext{}{\textbf{Note.} Units are in[$\times10^{-3}$ eV cm$^{-3}$].}
\tablenotetext{}{\textbf{Note. } Gould Belt includes Taurus, Orion, Perseus, and Ophiuchi; Radcliff Wave includes Taurus, Orion, and Perseus. Notice that the energy density values of Ophiuchi are a thousand times higher.}
\end{deluxetable*}

We compare the upper limits with our sensitivity as well as the cosmic-ray energy density measured by AMS. We used the measurements and fit function in \cite{ams} to calculate the blue shaded region in Figure \ref{fig:cruplims} (and Figure \ref{fig:uplimsCRs} in Appendix \ref{app:results}), which corresponds to the $\pm 2\sigma$ uncertainty region of the extrapolation. Table \ref{table:CRcredint} contains the observed and expected upper limits on the cosmic-ray energy density for each of the clouds and stacked analyses.

\begin{figure*}[!htb]
\centering
\gridline{\fig{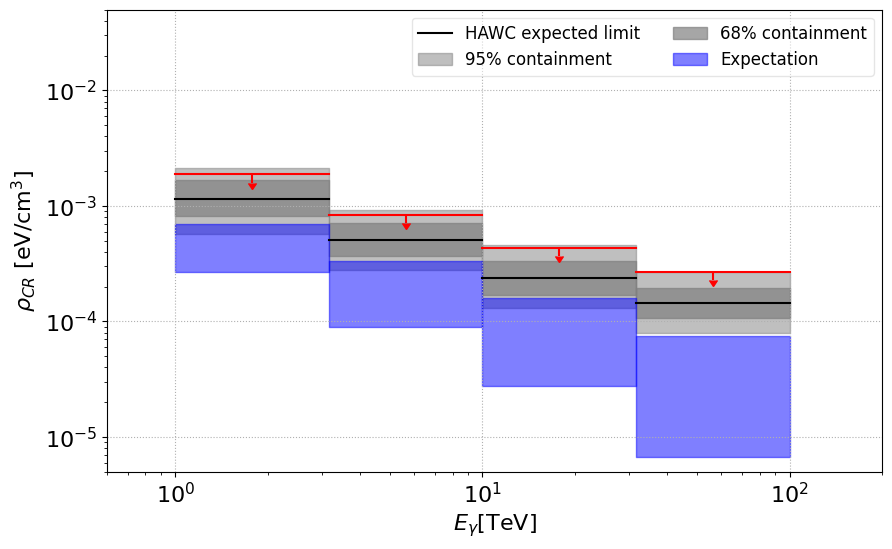}{0.4\textwidth}{(a) All}
          \fig{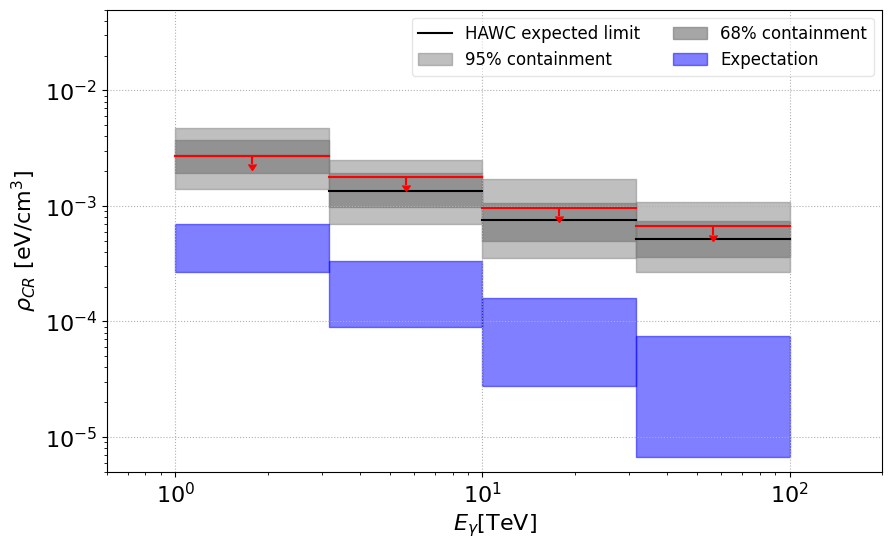}{0.4\textwidth}{(b) Gould Belt}}
\gridline{\fig{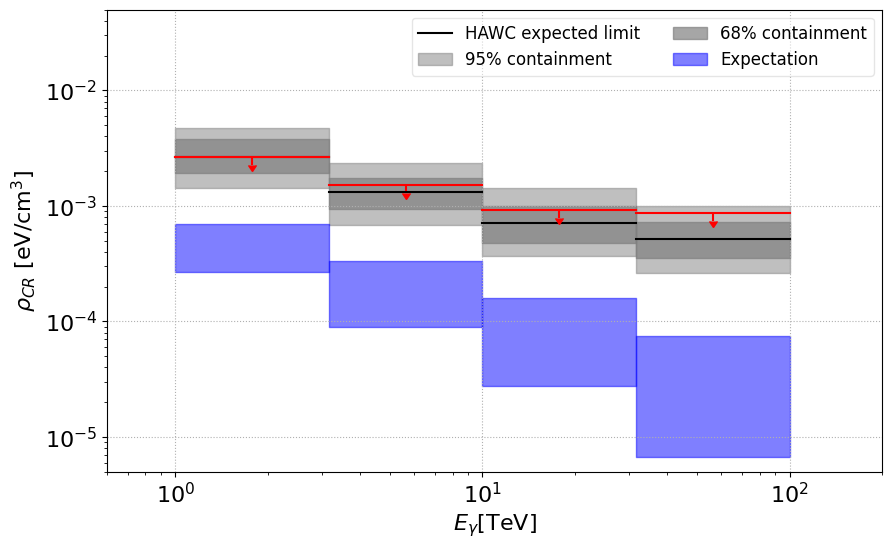}{0.4\textwidth}{(c) Radcliff Wave}}
\caption{95\% C.I. upper limits on the cosmic-ray energy density from the stacked analyses (All the clouds, Gould Belt and Radcliff Wave structures) as a function of gamma-ray energy. The corresponding cosmic-energy density can be estimated based on the assumption that $E_{CR}\sim10\,E_{\gamma}$~\citep[see for example][]{crossSec2}. Gray band represents the statistical uncertainty in the U.L. (68$\%$ and 90$\%$ containment). Blue band corresponds to the the $\pm 2\sigma$ uncertainty region  of the cosmic ray energy density using data (and extrapolating) from AMS. }
\label{fig:cruplims}
\end{figure*}

From these limits we do not have evidence that there is any deviation from the paradigm of the sea of cosmic rays permeating our Galaxy. It is also important to remark again that the gamma-ray expectation is based on an extrapolation of cosmic-ray data, so these estimated limits give a constraint on the cosmic ray energy density at energies of $E_{CR}\sim10\,E_{\gamma}$~\citep[see for example][]{crossSec2}, assuming purely pion decay photons.

\section{Conclusions}
\label{sec:Conclusion}

With the purpose to test the cosmic-ray sea paradigm and probe if its density is independent of the location, we performed measurements of gamma-ray flux from high-latitude passive GMCs using data from the HAWC observatory. Since no significant excess was observed, we calculated upper limits at the 95$\%$ credible interval, for individual and stacked gamma-ray emission of clouds that are part of the Gould Belt as described in $\S$\ref{sec:clouds}. The gamma-ray flux expected from pure hadronic interactions of the cosmic-ray flux with passive molecular clouds is below $10^{-11}$ TeV$^{-1}$ cm$^{-2}$ s$^{-1}$ above 10\,TeV. The quasi-differential limits are less than a factor of 10 higher than the prediction.
The most stringent constraint that we observe is in the 2.7-model limit when stacking all clouds. However, with the current data, we do not find enough evidence to reject the paradigm of the sea of cosmic rays.

Using the same analysis method for the differential gamma-ray limits, we also estimated the cosmic-ray energy density of the clouds. Using the data from AMS and extrapolating the results to HAWC energies, we see that the HAWC limits are higher by a factor of less than 10 for the case of the stacked clouds as wells as Taurus and Aquila. 

With current detector settings, as described in $\S$\ref{sec:Detector}, doubling the exposure time would lower the constraints by a factor of 0.7, which will allow to make a definitive statement about the cosmic-ray flux in distant regions of the Galaxy, at least in the case of the stacked analysis. The sensitivity of this analysis will also improve with the help of new installed outrigger detectors at the HAWC site, as well as the current development of new algorithms for event reconstruction, gamma-hadron separation, and background estimation.

\acknowledgements
We acknowledge the support from: the US National Science Foundation (NSF); the US Department of Energy Office of High-Energy Physics; the Laboratory Directed Research and Development (LDRD) program of Los Alamos National Laboratory; Consejo Nacional de Ciencia y Tecnolog\'ia (CONACyT), M\'exico, grants 271051, 232656, 260378, 179588, 254964, 258865, 243290, 132197, A1-S-46288, A1-S-22784, c\'atedras 873, 1563, 341, 323, Red HAWC, M\'exico; DGAPA-UNAM grants IG101320, IN111315, IN111716-3, IN111419, IA102019, IN110621; VIEP-BUAP; PIFI 2012, 2013, PROFOCIE 2014, 2015; the University of Wisconsin Alumni Research Foundation; the Institute of Geophysics, Planetary Physics, and Signatures at Los Alamos National Laboratory; Polish Science Centre grant, DEC-2017/27/B/ST9/02272; Coordinaci\'on de la Investigaci\'on Cient\'ifica de la Universidad Michoacana; Royal Society - Newton Advanced Fellowship 180385; Generalitat Valenciana, grant CIDEGENT/2018/034; Chulalongkorn University’s CUniverse (CUAASC) grant. Thanks to Scott Delay, Luciano D\'iaz and Eduardo Murrieta for technical support.

\bibliography{hawcGMC} 

\appendix
\section{Mass calculation}\label{app:masses}
Although some of the mass values of the clouds can be found in the literature, we calculate them ourselves due to the fact that we are building the regions of interest for the analysis, focusing on the high-density areas of the clouds where we assume most of the gamma rays originate. 
As mentioned in the main text, we use data from the Planck survey. 
We start by calculating the column density:
\begin{equation}\label{eq:planckCD}
N_{H} = \tau_D / \left(\frac{\tau_D}{N_{H}}\right)^{ref}
\end{equation}
where the reference value used is $(\tau_D/N_{H})^{ref} = (1.18\pm0.17)\times10^{-26} \text{cm}^2$ for 353GHz \cite{planck}. $\tau_D$ is the measured dust opacity.
The mass of the cloud is then calculated as 
\begin{equation}
M_{dust} = N_{H} \Omega d^2 m_{H},
\end{equation}
where $\Omega$ is the angular area of the cloud and $d$ is the distance to the cloud.

We apply a cut on the dust opacity value to select the high-density regions of $5\times10^{-5}$ (for Hercules, in which the density is lower, we use a cut of $2.5\times10^{-5}$) as done in \cite{fermiCR}.

\section{Templates}\label{app:templates}

\begin{figure*}[!htb]
\gridline{\fig{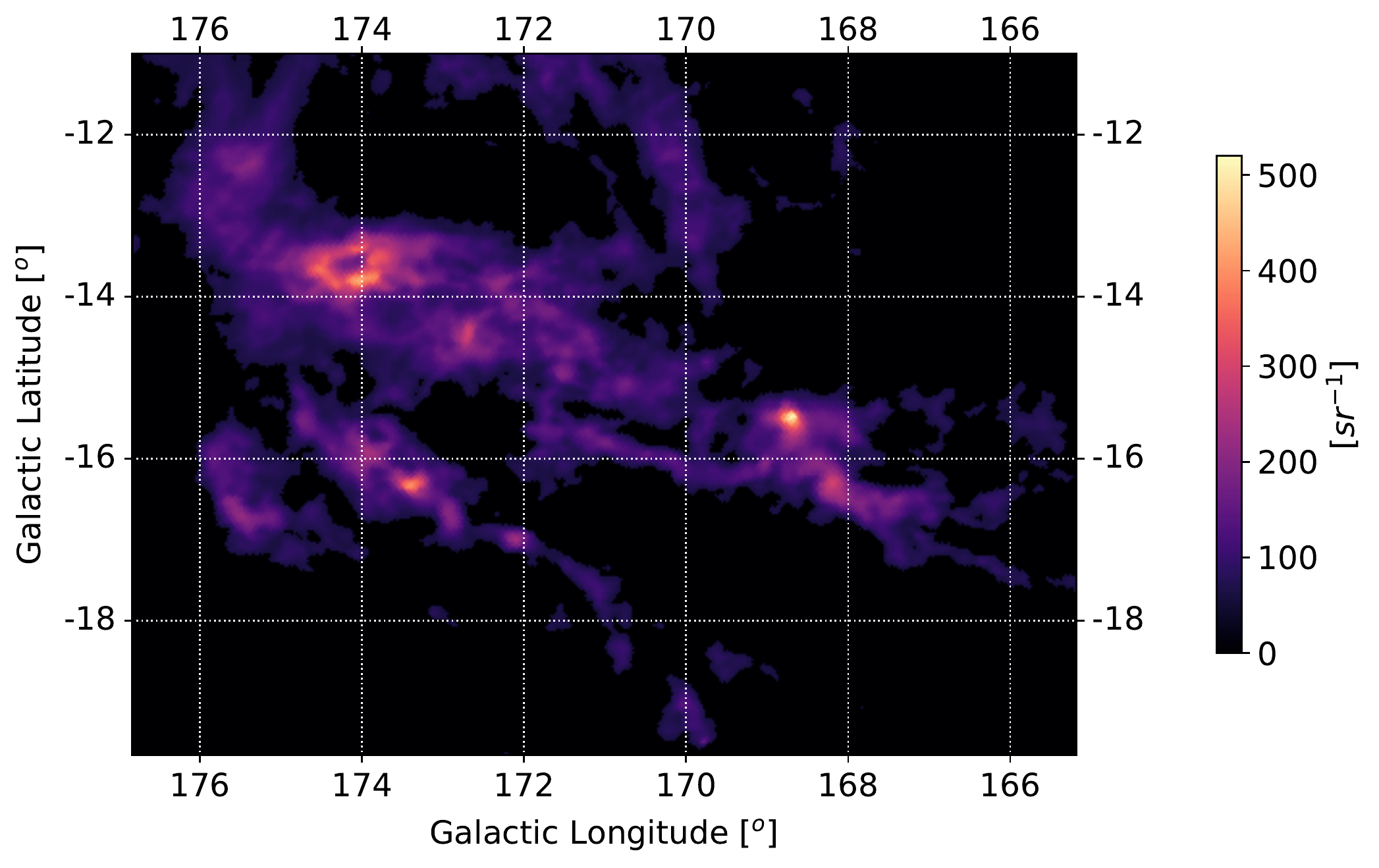}{0.4\textwidth}{(a) Taurus}
          \fig{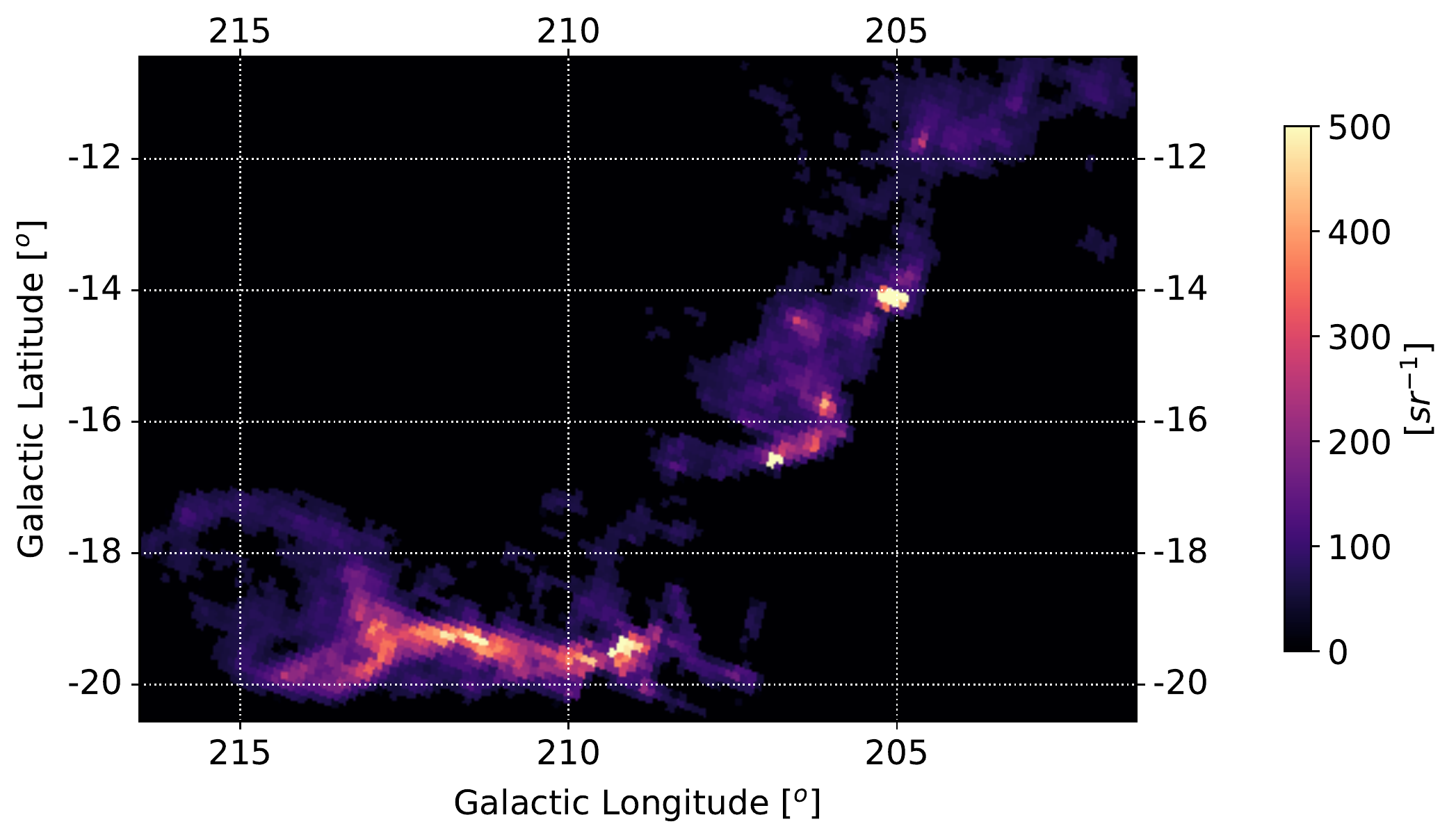}{0.4\textwidth}{(b) Orion}}
\gridline{\fig{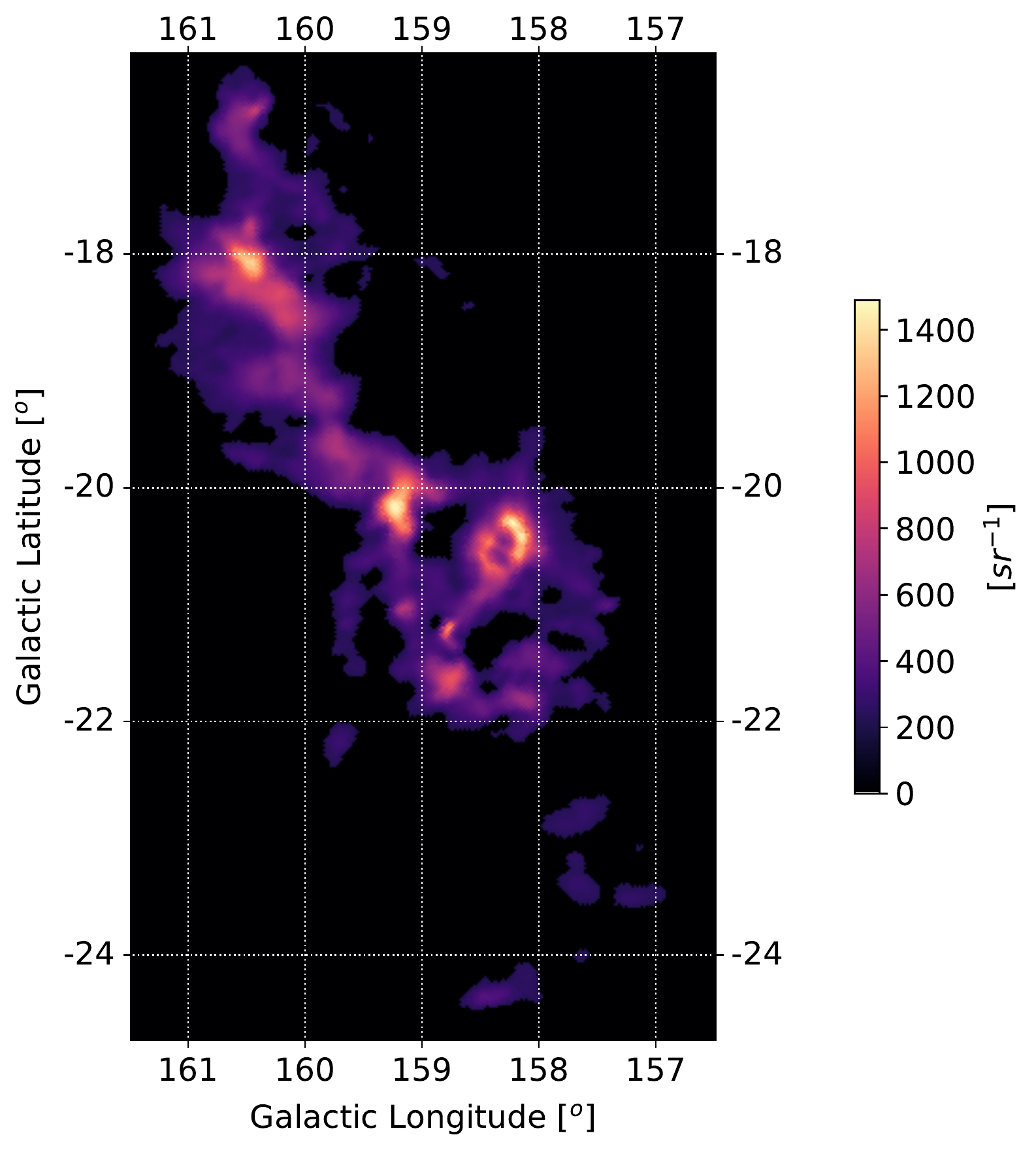}{0.3\textwidth}{(c) Perseus}
          \fig{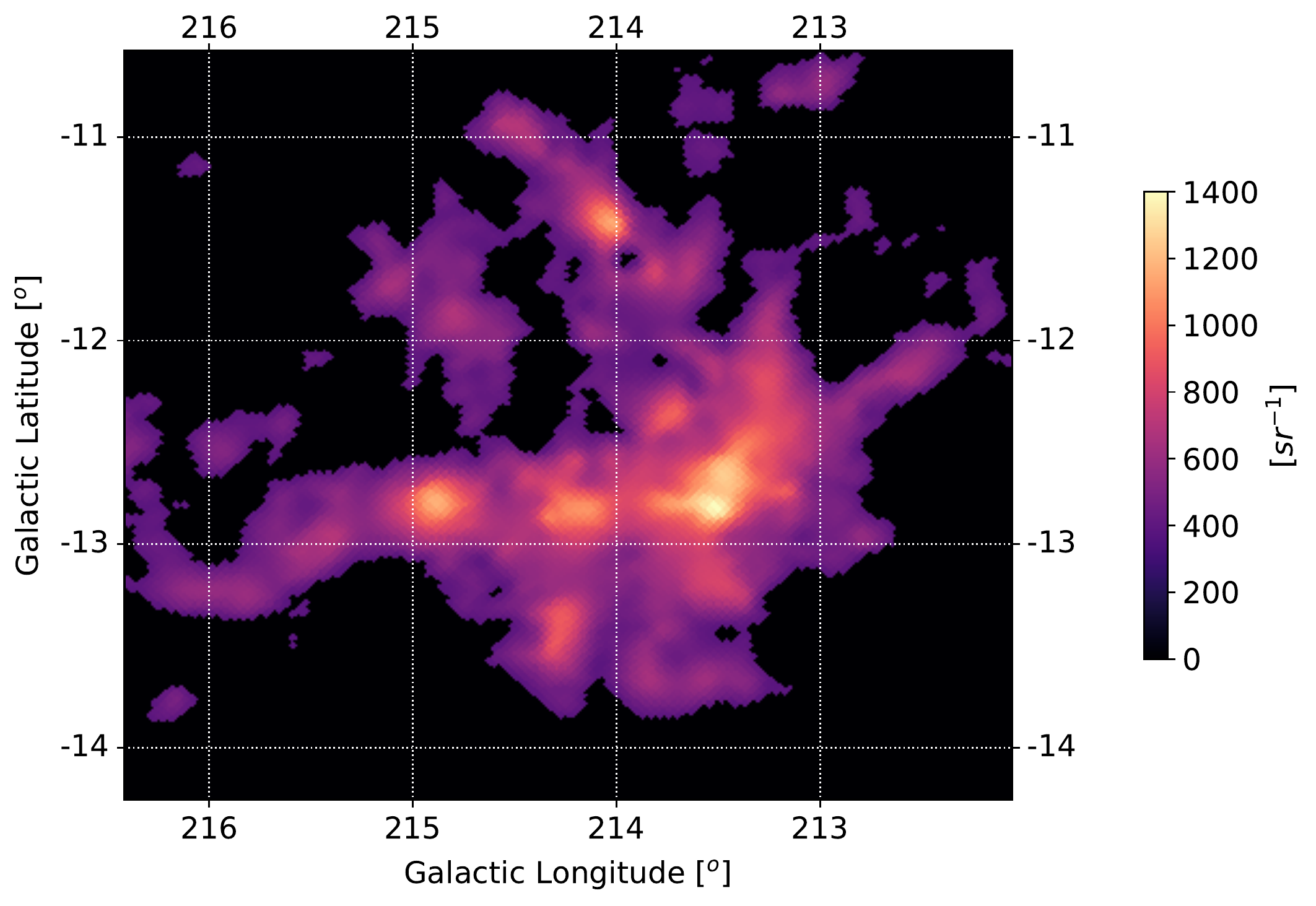}{0.4\textwidth}{(d) Monoceros}}
\gridline{\fig{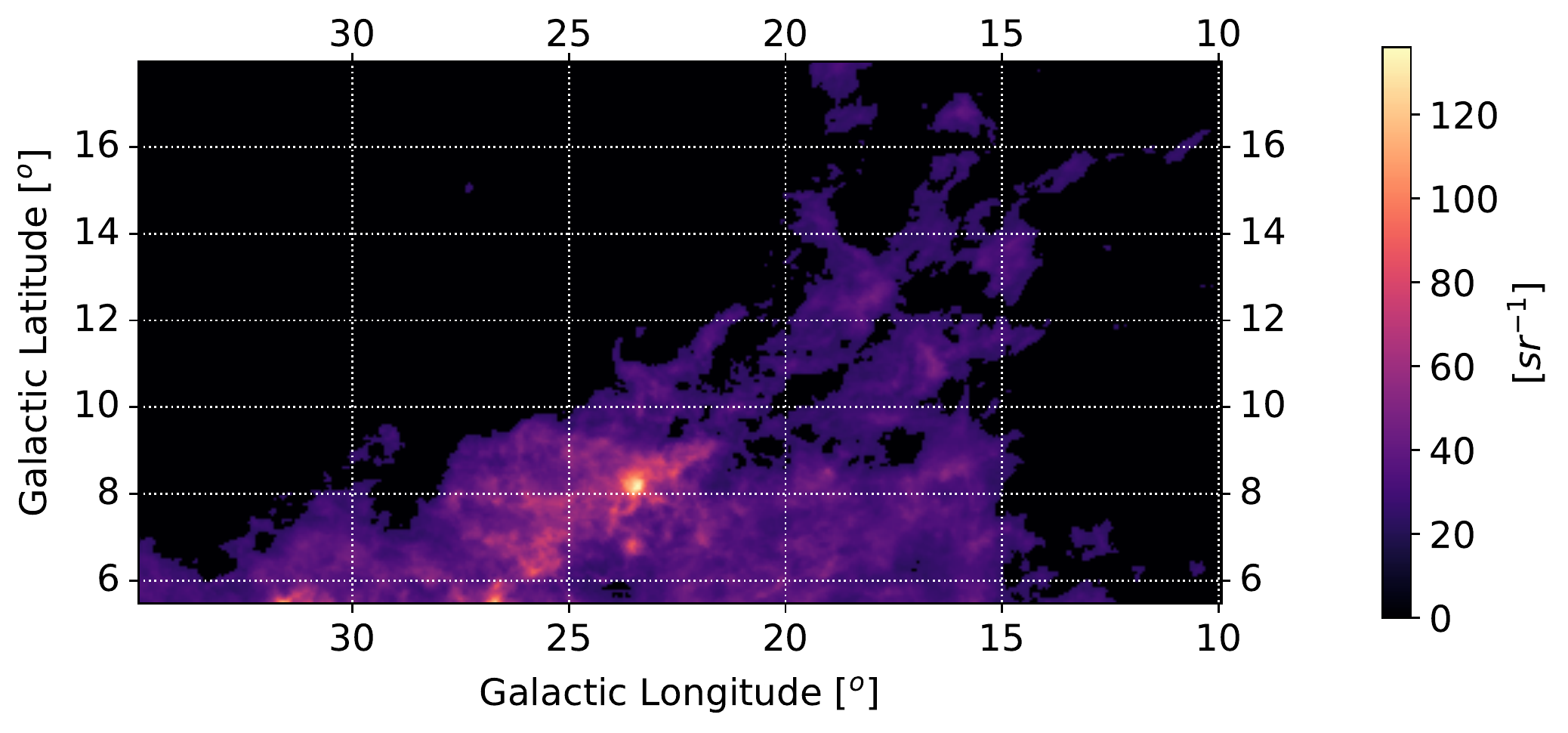}{0.4\textwidth}{(e) Aquila}
          \fig{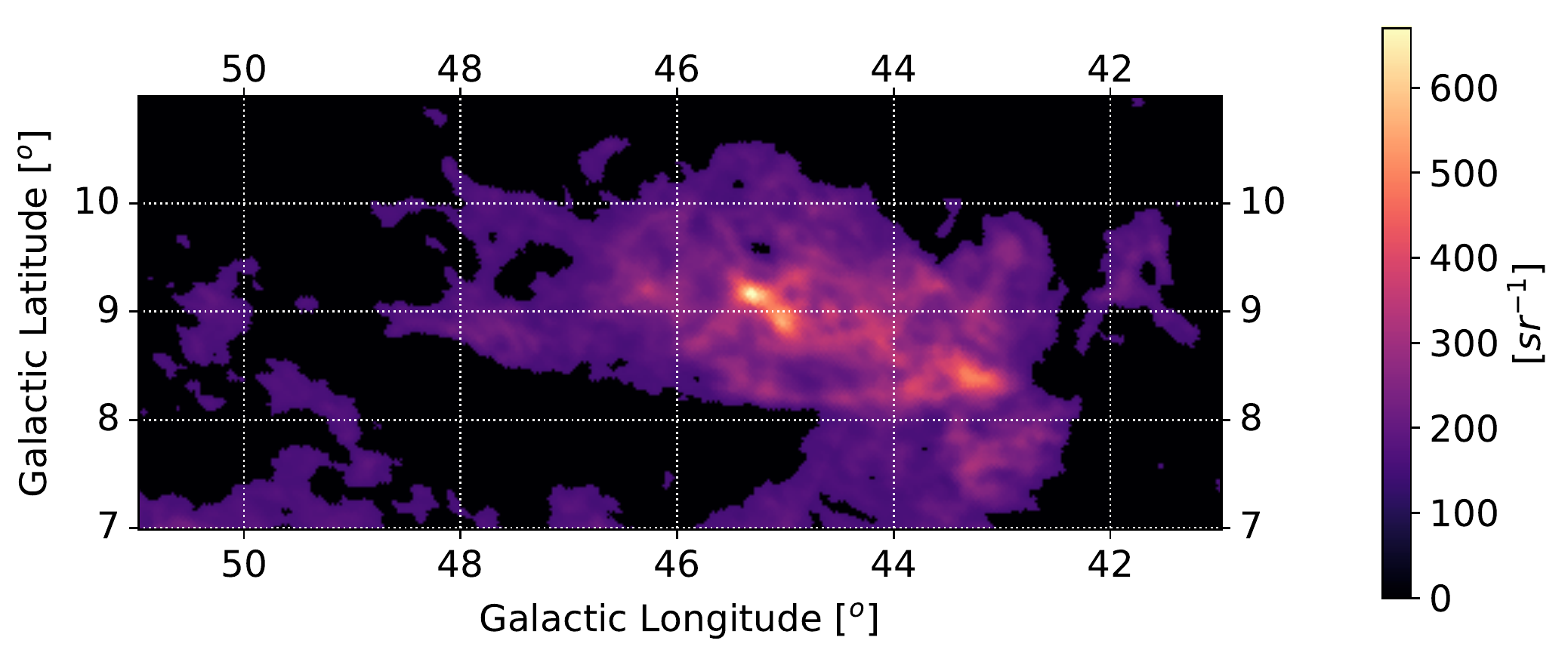}{0.4\textwidth}{(f) Hercules}} 
\gridline{\fig{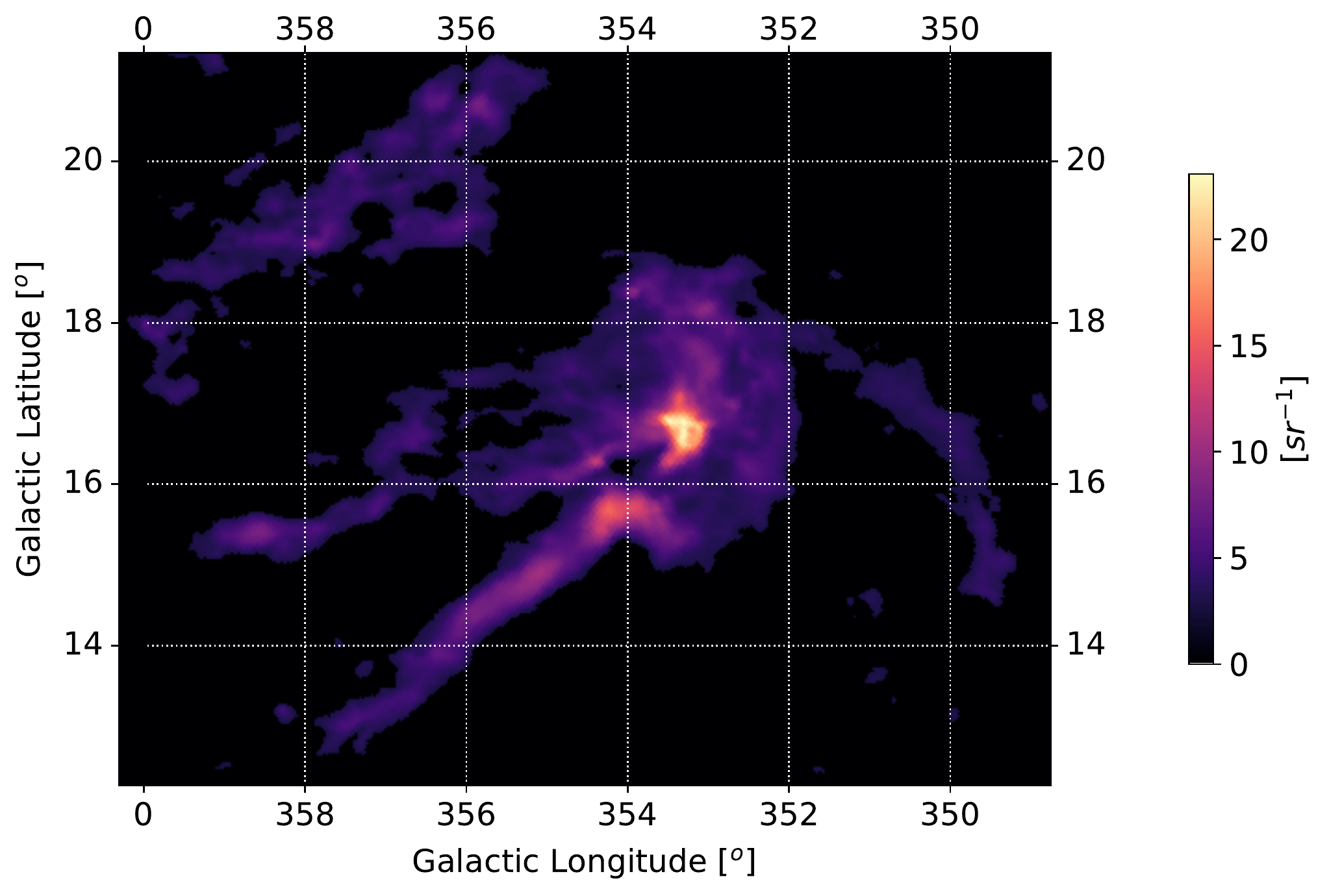}{0.4\textwidth}{(g) Ophiuchi}}
\caption{Templates created using the information from the Planck Survey. Colorbar is the normalized column density divided by the size of the spatial bin. }
\label{fig:template}
\end{figure*}

We generated our molecular cloud templates using the Planck survey. We apply the same methodolgy as in \cite{fermiCR}.  First we calculate the column density from the dust optical depth map at 353~GHz from Equation \ref{eq:planckCD}. We apply the corresponding cut on the opacity value as mentioned in Appendix \ref{app:masses}, then we normalize the map to the integral of the column density divided by the size of the spatial bin. This procedure ensures that the correct units are derived when the 3ML spatial model is used. Figure \ref{fig:template} shows the results after applying this procedure.

\section{Results of the Analysis}\label{app:results}

\begin{figure*}[!htb]
\centering
\gridline{\fig{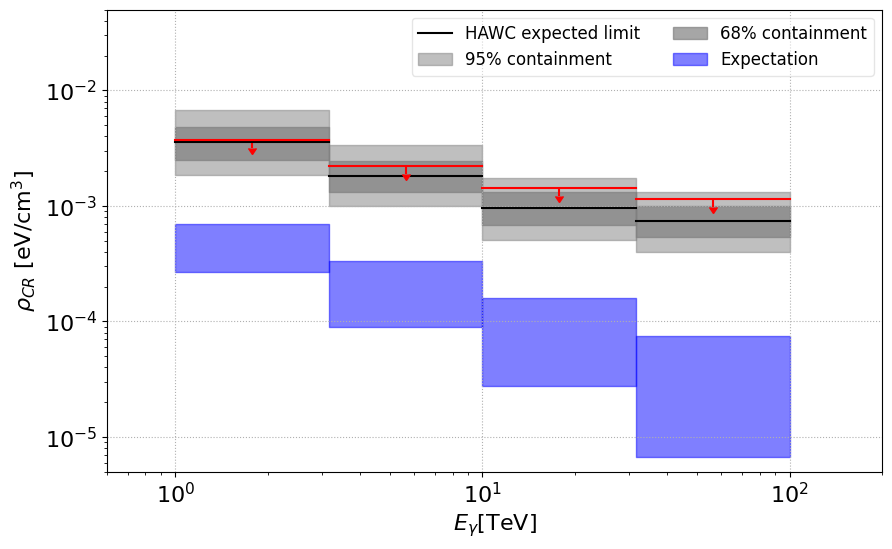}{0.4\textwidth}{(a) Taurus}
          \fig{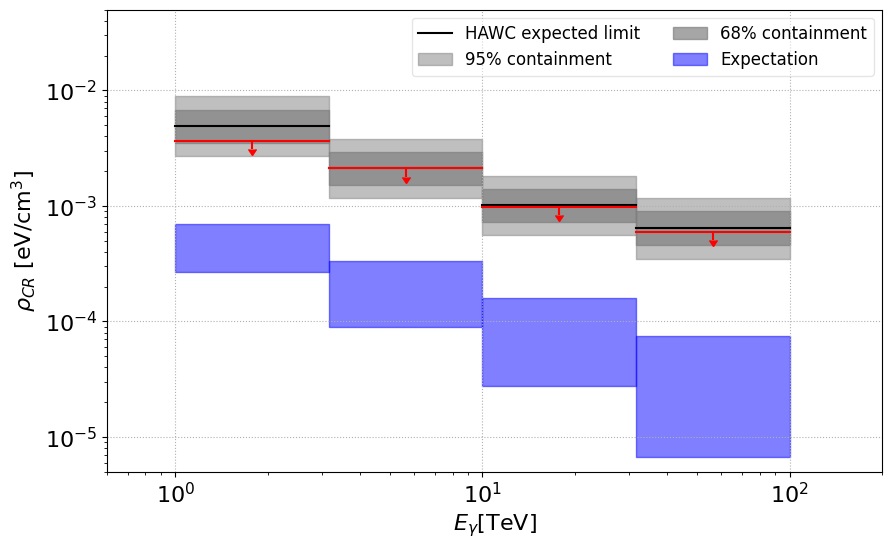}{0.4\textwidth}{(b) Orion}}
\gridline{\fig{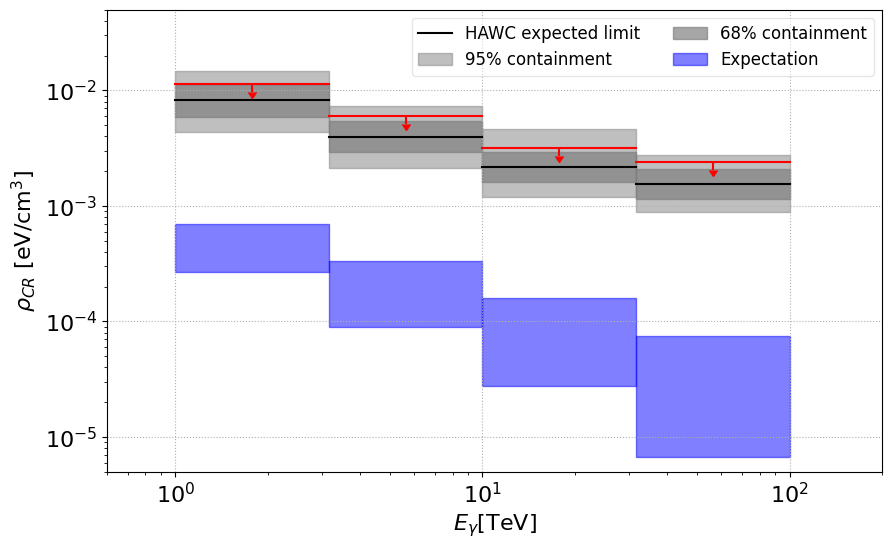}{0.4\textwidth}{(c) Perseus}
          \fig{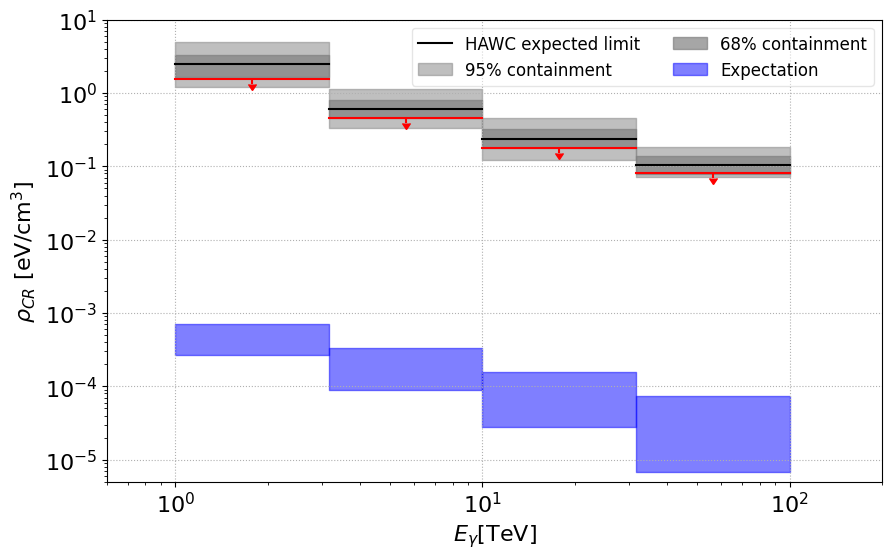}{0.4\textwidth}{(d) Ophiuchi}}
\gridline{\fig{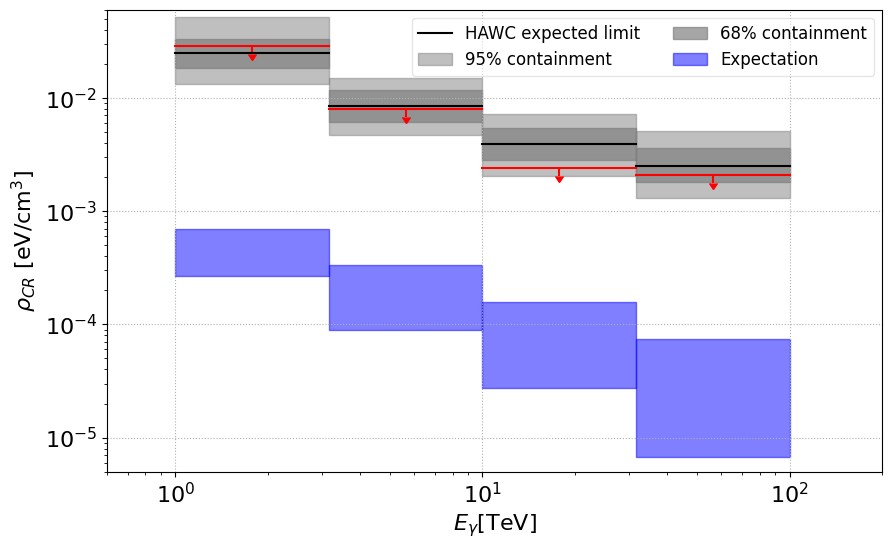}{0.4\textwidth}{(e) Monoceros}
          \fig{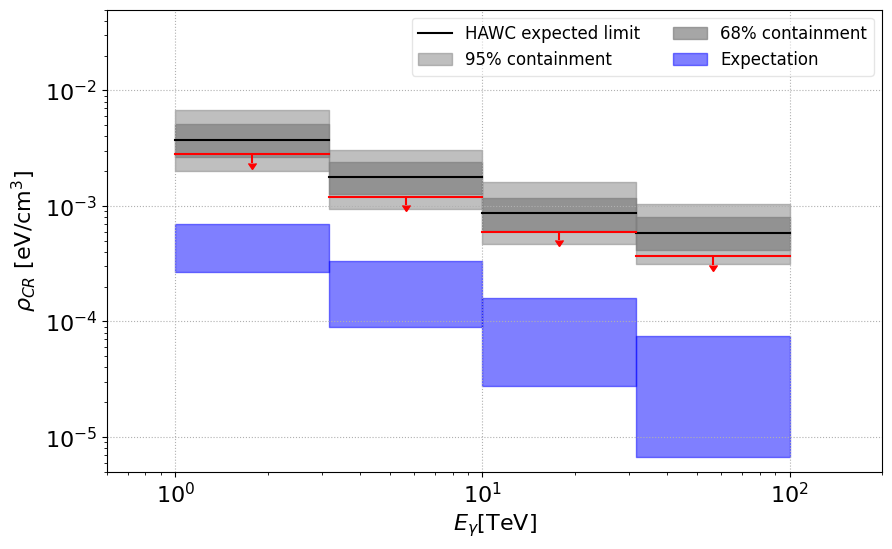}{0.4\textwidth}{(f) Aquila}}         
\gridline{\fig{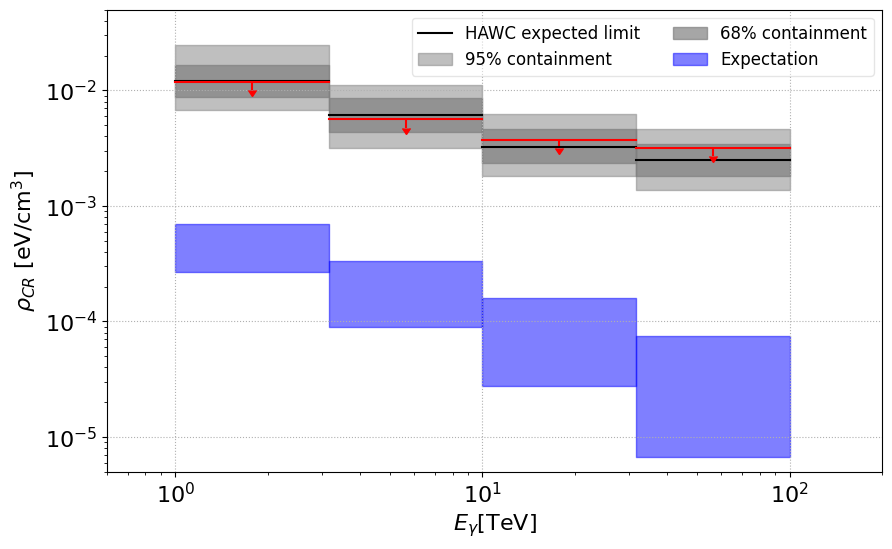}{0.4\textwidth}{(g) Hercules}}
\caption{Cosmic-ray density upper limits of the GMCs.}
\label{fig:uplimsCRs}
\end{figure*}

\begin{figure*}%
\figurenum{6}
\gridline{\fig{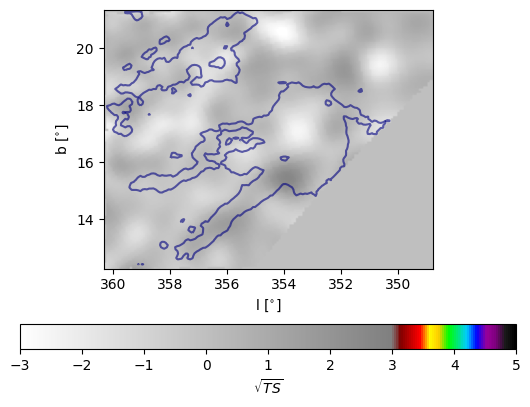}{0.35\textwidth}{(d) Ophiuchi}
          \fig{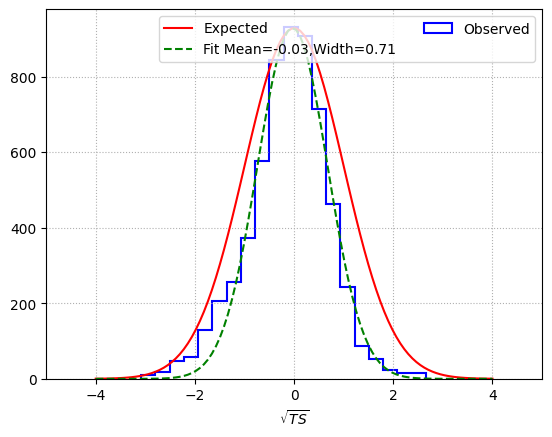}{0.35\textwidth}{}}
\gridline{\fig{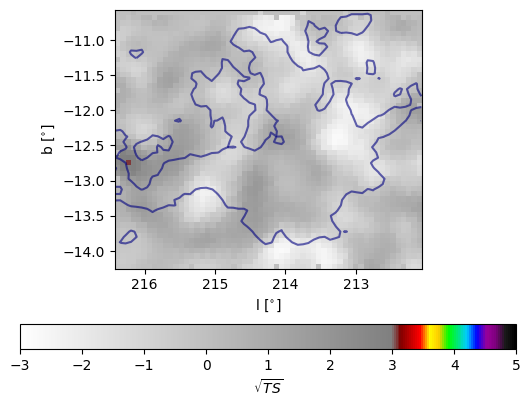}{0.35\textwidth}{(e) Monoceros}
          \fig{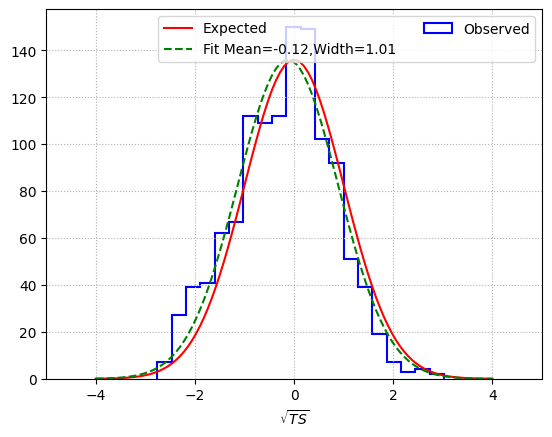}{0.35\textwidth}{}}
\gridline{\fig{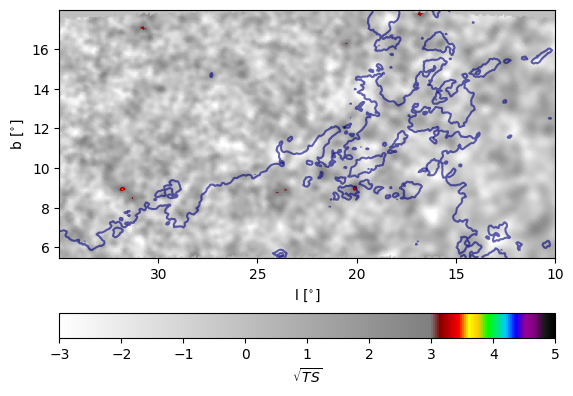}{0.35\textwidth}{(f) Aquila}
          \fig{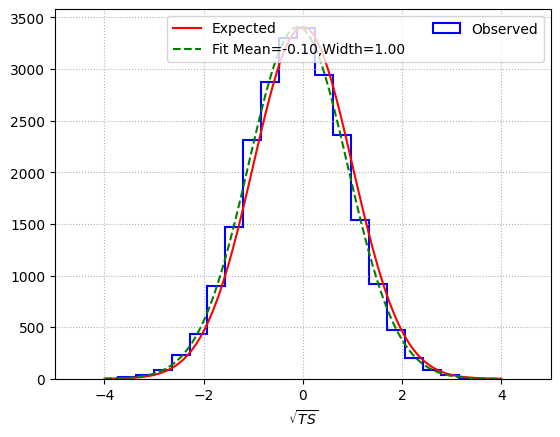}{0.35\textwidth}{}}        
\caption{HAWC significance maps of the GMC regions and their significance distribution. The blue contour line comes from the Planck Survey with the dust opacity value of $5\times10^{-5}$. The limit of the HAWC field of view can be seen in the Ophiuchi region (Continued).}
\label{fig:sigmaps}
\end{figure*}

Table \ref{table:credintplus} shows the TS results, together with the corresponding flux measurements of Table \ref{table:credint}. As can be seen, none of the results are significant (i.e. TS$>$25).
Similarly, Table \ref{table:CRcredintplus} shows the TS results and cosmic-ray energy density measurements of Table \ref{table:CRcredint}.
Figure \ref{fig:uplimsCRs} shows the results of the cosmic-ray energy density of the rest of the clouds that were not presented in the main text. 
Finally, Figure \ref{fig:sigmaps} shows the HAWC significance maps of the GMC regions as well as the distribution of the significance of the regions. We note that, except for two clouds, the distributions of the significance behave as expected: a Gaussian distribution with mean of 0 and width of 1. The Perseus region has a shifted mean of 0.27, while the Ophiuchi region has a width of 0.71. This could be related due to systematic effects in the background estimation at this particular declination or statistical fluctuations. To check this we rotated the two affected regions of interests in right ascension and confirmed that the significance distributions behaved as expected. The systematic uncertainties due to the deviation in the width and mean from the Gaussian expectation are not affecting the expected limits reported in $\S$\ref{sec:results}, for which we use Poisson-fluctuated background-only maps.

\begin{longrotatetable}
\begin{deluxetable*}{ccccccccccc}
\tablecaption{Gamma-ray flux measurements and 95\% C.I. upper limits (in parenthesis).\label{table:credintplus}}
\centering
\tablewidth{900pt}
\tabletypesize{\scriptsize}
\tablehead{
 \colhead{Energy Range} & \multicolumn{2}{c}{ 1 - 3.16 TeV } & \multicolumn{2}{c}{ 3.16 - 10.0 TeV } & \multicolumn{2}{c}{ 10.0 - 31.6 TeV } & \multicolumn{2}{c}{$>$31.6 TeV } & \multicolumn{2}{c}{ $>$1 TeV}\\
 \colhead{Pivot Energy}&\multicolumn{2}{c}{1.78 TeV}&\multicolumn{2}{c}{5.62 TeV} & \multicolumn{2}{c}{17.8 TeV} & \multicolumn{2}{c}{56.2 TeV} & \multicolumn{2}{c}{10 TeV}\\
}
\startdata
  & TS & Flux [$\times10^{-12}$] & TS & Flux[$\times10^{-13}$] & TS & Flux[$\times10^{-15}$] & TS & Flux[$\times10^{-16}$] & TS & Flux[$\times10^{-15}$]\\
 \textbf{Gould Belt} & 0.0 & 1.3$^{+1.4}_{-0.9}$(3.6) & 0.8 & 1.0$^{+0.8}_{-0.6}$(2.3) & 1.4 & 6.5$^{+4.2}_{-3.8}$(13.5) & 1.5 & 4.2$^{+2.9}_{-2.6}$(9.4) & 0.79 & 7.5$^{+5.9}_{-4.9}$(18.0) \\
  \textbf{Radcliff} & 0.0 & 1.0$^{+1.1}_{-0.7}$(2.8) & 0.8 & 0.8$^{+0.6}_{-0.5}$(1.8) & 1.4 & 4.6$^{+3.3}_{-2.8}$(10.3) & 1.5 & 3.4$^{+2.2}_{-1.9}$(7.4) & 0.79 & 5.9$^{+4.5}_{-3.8}$(13.8) \\
 \textbf{Taurus} & 0.0 & 0.6$^{+0.6}_{-0.4}$(1.5) & 0.5 & 0.4$^{+0.3}_{-0.3}$(1.0) & 1.5 & 2.9$^{+2.0}_{-1.7}$(6.4)& 2.0 & 2.5$^{+1.5}_{-1.4}$(5.0) & 0.8 & 3.4$^{+2.6}_{-2.2}$(7.8) \\
 \textbf{Orion} & 0.0 & 0.6$^{+0.8}_{-0.5}$(2.0) & 0.0 & 0.1$^{+0.4}_{-0.3}$(1.0) & 0.0 & 1.5$^{+1.9}_{-1.0}$(4.8) & 0.0 & 1.0$^{+1.2}_{-7.4}$(3.0) & 0.0 & 2.3$^{+2.6}_{-1.6}$(7.1)\\
 \textbf{Perseus} & 1.3 & 0.6$^{+0.5}_{-0.4}$(1.4) & 1.9 & 0.4$^{+0.2}_{-0.2}$(0.74) & 2.3 & 2.0$^{+1.2}_{-1.0}$(4.0) & 2.3 & 1.5$^{+0.9}_{-0.8}$(2.9) & 2.8 & 3.0$^{+1.7}_{-1.6}$(5.8)\\
 \textbf{Ophiuchi [x100]} & 0.0 & 1.4$^{+1.8}_{-1.0}$(4.7) & 0.0 & 0.4$^{+0.6}_{-0.3}$(1.4) & 0.0 & 1.6$^{+2.0}_{-1.1}$(5.3) & 0.0 & 0.8$^{+1.0}_{-0.5}$(2.5) & 0.0 & 2.7$^{+3.5}_{-2.0}$(8.6)\\
 \textbf{Monoceros} & 0.5 & 0.8$^{+0.7}_{-0.5}$(1.9) & 0.0 & 0.2$^{+0.2}_{-0.1}$(0.52) & 0.0 & 0.4$^{+0.7}_{-0.3}$(1.7) & 0.0 & 0.8$^{+3.0}_{-0.7}$(1.2) & 0.0 & 1.0$^{+1.3}_{-0.7}$(3.4) \\
\textbf{Aquila} & 0.0 & 1.0$^{+1.1}_{-0.7}$(3.2) & 0.0 & 0.4$^{+0.5}_{-0.3}$(1.3) & 0.0 & 1.9$^{+2.7}_{-1.4}$(6.9) & 0.0 & 1.2$^{+1.6}_{-0.8}$(4.2) & 0.0 & 2.6$^{+3.7}_{-1.9}$(9.8)\\
 \textbf{Hercules} & 0.0 & 0.4$^{+0.4}_{-0.2}$(1.0) & 0.0 & 0.2$^{+0.2}_{-0.1}$(0.5) & 0.0 & 1.1$^{+1.1}_{-0.8}$(3.2) & 0.44 & 1.2$^{+0.9}_{-0.8}$(2.9) & 0.0 & 1.5$^{+1.6}_{-1.0}$(4.2) \\
\enddata
\tablenotetext{}{\textbf{Note. } Flux units are in TeV$^{-1}$ cm$^{-2}$ s$^{-1}$.}
\tablenotetext{}{\textbf{Note. } Gould Belt includes Taurus, Orion, Perseus, and Ophiuchi; Radcliff Wave includes Taurus, Orion, and Perseus}
\end{deluxetable*}
\end{longrotatetable}

\begin{longrotatetable}
\begin{deluxetable*}{ccccccccc}
\tablewidth{6cm}
\tabletypesize{\scriptsize}
\tablecaption{Cosmic-ray energy density quasi-differential 95\% C.I. upper limits  [$\times10^{-3}$ eV cm$^{-3}$] in each energy bin.\label{table:CRcredintplus}}
\tablehead{
 \colhead{Energy Range} & \multicolumn{2}{c}{ 1 - 3.16 TeV } & \multicolumn{2}{c}{ 3.16 - 10.0 TeV } & \multicolumn{2}{c}{ 10.0 - 31.6 TeV } & \multicolumn{2}{c}{$>$31.6 TeV } \\
 \colhead{Pivot Energy} &\multicolumn{2}{c}{1.78 TeV}&\multicolumn{2}{c}{5.62 TeV} & \multicolumn{2}{c}{17.8 TeV} & \multicolumn{2}{c}{56.2 TeV} \\
}
\startdata
  & TS &  & TS &  & TS &  & TS &  \\
\textbf{All} & 0.0 & 0.7$^{+0.6}_{-0.5}$(1.9) & 0.0 & 0.4$\pm0.3$(0.8) & 0.0 & 0.2$\pm0.2$(0.4) & 0.0 & 0.1$\pm0.1$(0.2)  \\
\textbf{Gould Belt} & 0.0 & 1.1$^{+0.9}_{-0.8}$(2.7) & 0.9 & 0.8$\pm0.5$(1.8) & 1.4 & 0.5$\pm0.3$(1.0) & 1.4 & 0.3$\pm0.2$(0.7)  \\
\textbf{Radcliff} & 0.0 & 1.1$\pm0.8$(2.7) & 0.9 & 0.8$\pm0.5$(1.5) & 1.5 & 0.5$\pm0.3$(0.9) & 1.4 & 0.3$\pm0.2$(0.9) \\
\textbf{Taurus} & 0.0 & 1.5$\pm1.1$(3.6) & 0.5 & 0.9$\pm0.7$(2.2) & 1.5 & 0.7$^{+0.4}_{-0.3}$(1.4) & 2.0 & 0.6${^{+0.4}_{-0.3}}$(1.2)  \\
\textbf{Orion} & 0.0 & 1.4$\pm1.1$(3.7) & 0.0 & 0.8$\pm0.6$(2.0) & 0.0 & 0.4$\pm0.7$(0.9) & 0.0 & 0.2$\pm0.2$(0.6)  \\
\textbf{Perseus} & 1.3 & 5.6$^{+3.4}_{-3.5}$(11.3) & 1.9 & 3.1$^{+1.7}_{-1.8}$(6.0) & 2.3 & 1.7$\pm0.9$(3.4) & 2.3 & 1.2$\pm0.7$(2.4)  \\
\textbf{Ophiuchi [x1000]} & 0.0 & 0.6$\pm0.5$(1.6) & 0.0 &  0.2$\pm0.1$(0.5) & 0.0 & 0.07$^{+0.06}_{-0.05}$(0.2) & 0.0 & 0.03$\pm0.02$(0.08)  \\
\textbf{Monoceros} & 0.5 & 13.0$\pm9.0$(28.6) & 0.0 &  3.1$\pm2.4$(7.9) & 0.0 & 0.9$\pm0.7$(2.4) & 0.0 & 0.7$\pm0.6$(2.0)  \\
\textbf{Aquila} & 0.0 & 1.1$\pm0.8$(2.8) & 0.0 & 0.4$^{+0.4}_{0.3}$(1.2) & 0.0 & 0.2$\pm0.2$(0.6) & 0.0 & 0.1$\pm0.1$(0.4)  \\
\textbf{Hercules} & 0.0 & 5.0$\pm4.0$(11.8) & 0.0 & 2.2$^{+1.7}_{-1.6}$(5.6) & 0.1 & 1.6$\pm1.1$(3.7) & 0.4 & 0.1 $\pm0.1$(3.2)  \\
\enddata
\end{deluxetable*}
\end{longrotatetable}

\begin{figure*}[!tb]%
\centering
\gridline{\fig{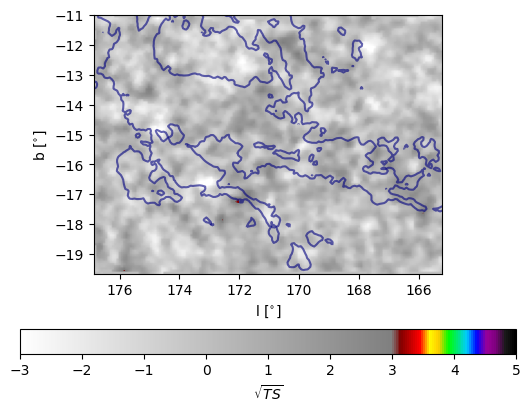}{0.35\textwidth}{(a) Taurus}
          \fig{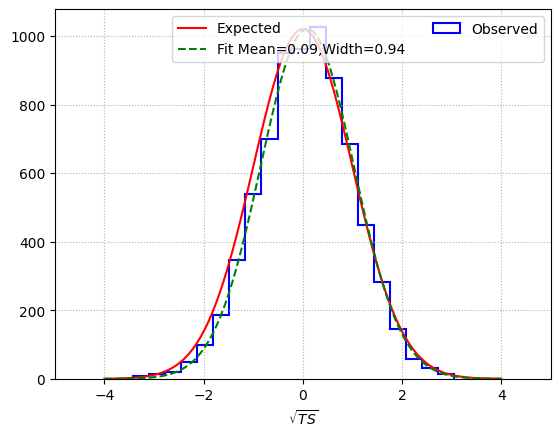}{0.35\textwidth}{}}
\gridline{\fig{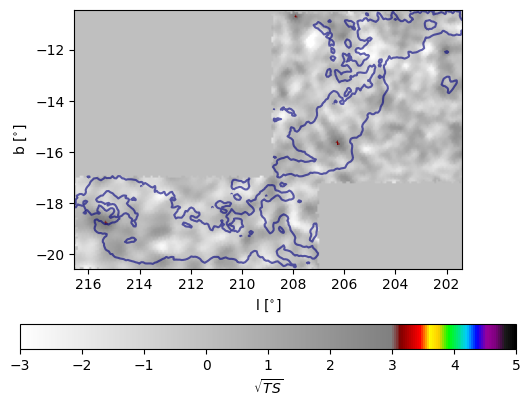}{0.35\textwidth}{(b) Orion}
          \fig{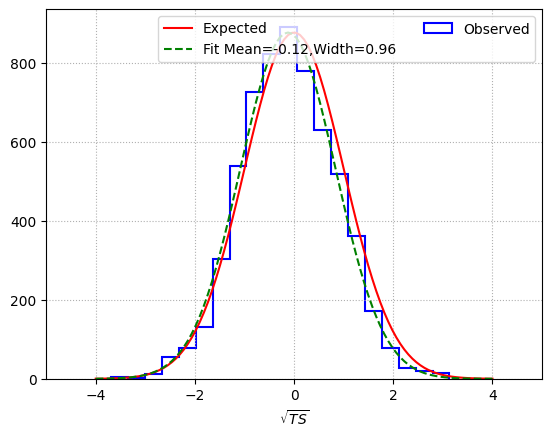}{0.35\textwidth}{}}
\gridline{\fig{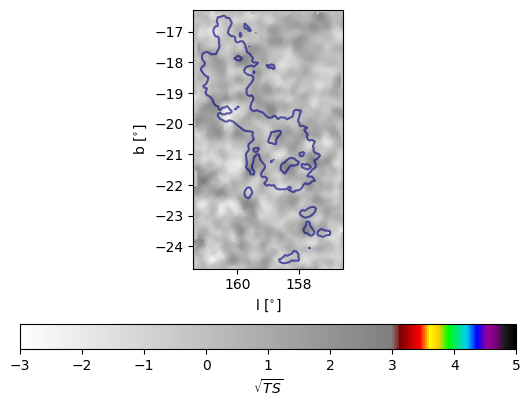}{0.35\textwidth}{(c) Perseus}
          \fig{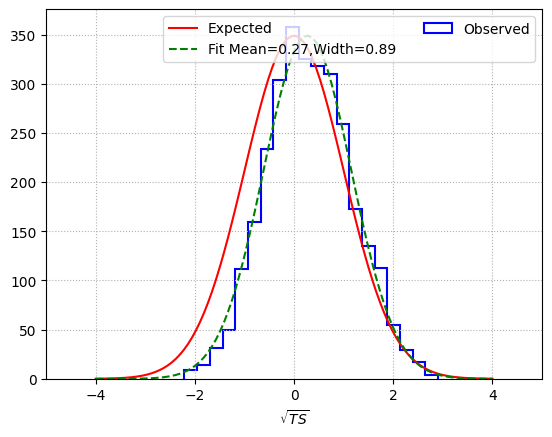}{0.35\textwidth}{}}
\caption{HAWC significance maps of the GMC regions and their significance distribution. The blue contour line comes from the Planck Survey with the dust opacity value of $5\times10^{-5}$. For Orion we only focus on the region of the cloud.}
\end{figure*}%
\begin{figure*}%
\figurenum{6}
\gridline{\fig{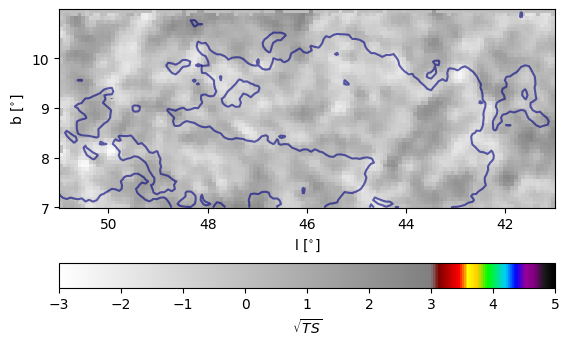}{0.35\textwidth}{(g) Hercules}
          \fig{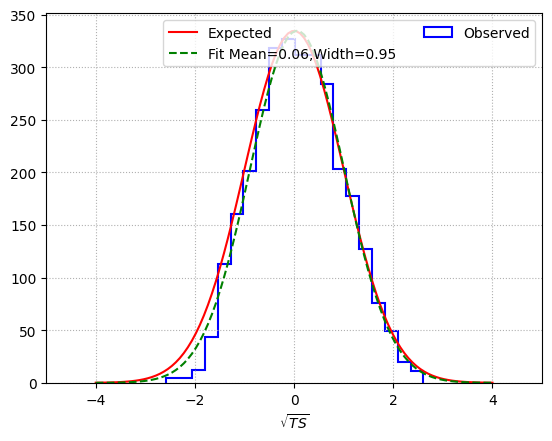}{0.35\textwidth}{}}
\caption{HAWC significance maps of the GMC regions and their significance distribution. The blue contour line comes from the Planck Survey with the dust opacity value of $2.5\times10^{-5}$ for Hercules. (Continued).}
\label{fig:skymaps}
\end{figure*}
\explain{New figures and caption were added in Figure: \ref{fig:sigmaps}}
\listofchanges
\end{document}